\documentclass{aa}  
\makeatletter
\providecommand{\columnwiselinenumbersfalse}{}
\providecommand{\columnwiselinenumberstrue}{}
\let\linenumbers\relax
\let\endlinenumbers\relax
\makeatother

\usepackage{graphicx}
\usepackage{txfonts}
\usepackage[compatibility=false]{caption}
\usepackage{subcaption}         
\usepackage{lscape}             
\usepackage{placeins}           
\usepackage{hyperref}
\usepackage{multirow}
\usepackage{float}
\usepackage{cleveref}
\usepackage{xcolor}
\usepackage{amsmath}
\usepackage{natbib}
\usepackage{stfloats}
\usepackage{cuted}
\usepackage{amsmath}
\begin{document}

   \title{AstroSat-UVIT observations of a possibly interacting pair of galaxies in HCG~77 }

   \subtitle{}

   \author{Anshul Srivastava\inst{1}\fnmsep\thanks{ph23032@iisermohali.ac.in}
   , H.K. Jassal\inst{1}\fnmsep\thanks{hkjassal@iisermohali.ac.in}, Mamta Gulati \inst{2} \and  K.P. Singh \inst{1,3}\fnmsep\thanks{kps@iisermohali.ac.in}
        }

   \institute{Indian Institute of Science Education 
   and Research, Mohali, SAS Nagar, Punjab 140306, India\\
            \and Thapar Institute of Engineering and Technology, Patiala, Punjab 147004, India\\ 
            \and Tata Institute of Fundamental Research, Mumbai 400005, India. }

   \date{}

  \abstract
   {Interactions and mergers possibly enhance the star formation in dwarf galaxies in a group environment.}
   { We aim to  study star-forming regions and the spectral energy distribution of two possibly interacting galaxies, PGC~56121 and PGC~56125, in the Hickson Compact Group 77.} 
   { We utilized the far-ultraviolet (FUV) channel of the Ultra Violet Imaging Telescope (UVIT) on board {\it AstroSat} to observe and produce FUV images of the galaxies. Our FUV images are at a much higher resolution in comparison to those obtained from previous galaxy surveys by GALEX in the near-UV and those from PS-1, DSS. }
   {We have identified several star-forming regions in the two possibly interacting galaxies, PGC~56121 and PGC~56125. These two galaxies form a pair widely separated in redshift and are seen in projection. We also report the presence of a candidate tidal dwarf galaxy at the end of one of the tidal tails located to the east of the pair, and we identified three major star-forming regions in the tidal dwarf. The spectral energy distribution of three galaxies in the system is presented and analyzed to investigate the key physical properties, such as stellar mass, dust mass, total luminosity, and star formation history, of the three galaxies. Based on these observations and on a comparison with observations in radio, these three galaxies are probably part of a small group of interacting galaxies.}
   {}

   \keywords{Galaxies: dwarf -- Galaxies: interactions -- Galaxies: groups: general -- Ultraviolet: general -- Techniques: image processing -- Galaxies: photometry
            }
   \authorrunning{Anshul Srivastava et al.}

   \maketitle

\section{Introduction}

Galaxy interactions represent one of the most dynamic and influential processes in the evolution of the Universe. These gravitational encounters, ranging from close flybys to full-scale mergers, play a pivotal role in shaping the structure, composition, and star formation activity of galaxies. From minor tidal disruptions that trigger localized starburst regions to cataclysmic mergers that forge giant elliptical galaxies, such interactions have been shown to significantly impact galactic growth and evolution (\citet{1972ApJ...178..623T}; \citet{1992ARA&A..30..705B}).

Galaxy interactions occur across all mass and length scales, from encounters between massive spirals and ellipticals to those involving the smallest dwarf systems. Dwarf galaxies are the most common galaxy type in the Universe, and they serve as crucial building blocks in hierarchical galaxy formation models. Their evolution is strongly shaped by both internal mechanisms, such as stellar feedback, and external interactions, such as tidal forces. Dwarf galaxies in pairs or small groups often exhibit elevated star formation rates (SFRs). Previous surveys found that interacting dwarf pairs can exhibit enhancements in SFR compared to their isolated counterparts (\citet{Stierwalt_2015}).
Interacting dwarfs show an increase in SFR relative to non-interacting galaxies in the stellar mass range $10^7 - 10^8 M_{\odot}$ (\citet{refId0}). 

While major mergers significantly enhance star formation, many morphological disturbances in dwarfs arise from minor interactions or flybys. Mergers moderately boost SFRs, but more frequent non-merging interactions cumulatively contribute significantly to stellar mass growth over cosmic time (\citet{Martin_2020}). 
Among the outcomes of dwarf–dwarf or dwarf–giant interactions, tidal dwarf galaxies (TDGs) are particularly intriguing, as they originate from preenriched material stripped from parent galaxies. Understanding the nature of these systems is crucial to assessing how tidal interactions shape the formation and evolution of dwarf galaxies either in groups or in compact groups.

An atlas of compact groups was provided by \citet{1993ApL&C..29....1H}. It contains a collection of galaxies designated as Hickson Compact Groups (HCGs),
   and has a list of one hundred groups.  HCG~77 is one of them (\citet{1982ApJ...255..382H}). In the original classification scheme, four galaxies were considered members of this group: PGC 51625, PGC~56121, UGC 10049, and PGC 56122. Later, using the redshift measurements, Hickson showed that the first two members have low redshifts very close to each other, and thus they either form a pair or are part of a single galaxy \citet{1993ApL&C..29....1H}.  The other two objects, which are much redder in color, have much larger redshifts, probably forming a completely different pair of galaxies, and they cannot be considered as belonging to HCG~77.  The group, therefore, has been misclassified as a compact group. However, observations with the Very Large Array show that the galaxies PGC~56121 and PGC~56125 may be part of a small local group, the most massive galaxy being UGC 10043 (\citet{Matthews_2004}; \citet{2009RMxAC..35..201A}).

 In this paper, we present far-ultraviolet (FUV) observations of HCG~77, where our focus is to study the pair of galaxies. 
This pair is located at a redshift of $z=0.006969 \pm 5.55E-6$ 
 and $z=0.007505 \pm 2.10E-4$ corresponding to a distance of $d \approx 29.80 \pm 0.02 \ Mpc$ and $d \approx 32.15 \pm 0.90 \ Mpc$ (with $H_{0}= 70 \ km/s/Mpc$) respectively.  
We performed aperture photometry of the star-forming complexes, and spectral energy distribution (SED) analysis of the system to obtain physical parameters such as the stellar mass, SFR, and specific SFR. 
The physical parameters are consistent with these galaxies being irregular and blue compact dwarfs. 
We also show the presence of a candidate TDG at the end of one of the tidal tails, whose size and color are consistent with the statistical study in \citet{Kaviraj_2011}. 

Observational studies and numerical simulations have shown that gravitational interactions can induce gas inflows, trigger starburst activity, and lead to the transformation of late-type spirals into early-type galaxies (\citet{1996ApJ...464..641M};\citet{Hopkins_2008}).  In addition to influencing star formation, galaxy interactions can also ignite nuclear activity. Tidal torques and dynamical friction can funnel gas toward the central regions of galaxies, feeding supermassive black holes and triggering active galactic nuclei activity(\citet{1989Natur.340..687H}; \citet{Hopkins_2006}).
The correlation between starburst and active galactic nuclei phases in interacting galaxies supports the idea of a coevolutionary link between black hole growth and galaxy evolution (\citet{Kormendy_2013}). Such processes are especially prominent in dense environments, such as compact groups or clusters, where close encounters and repeated interactions are more frequent (\citet{10.1111/j.1365-2966.2008.13531.x}; \citet{1982ApJ...255..382H}). By examining interacting systems across multiple wavelengths (e.g., ultraviolet, optical, infrared, and radio wavelengths), we can gain comprehensive insight into the physical mechanisms driving galaxy transformation in the evolving Universe.

The paper is organized as follows. In Sect. 2 we present the observational details. In Sect. 3 we discuss the image reduction pipeline used for the data. The final images, surface brightness contours, comparison with other images, identification, and photometry of the SFRs are presented in Sect. 4. We present the SED modeling and infer various physical parameters of the system in Sect. 5. Sect. 6 presents a discussion of our results, and Sect. 7 provides a summary of the study.

\section{Observation}
Observations of a field centered on HCG~77 were carried out by the Ultra-Violet Imaging Telescope (UVIT) (\citet{Tandon_2017}) on board the {\it AstroSat} (\citet{2014SPIE.9144E..1SS}).  The UVIT employs an $f/12$
Ritchey-Chretien telescope with an aperture of $375$ mm and has an intensified C-MOS imager at its focus. The observations were carried out on 18 February 2022, starting at 00:46:00, in the photon counting mode and using the $BaF_{2}$ filter with a central wavelength of 154 nm and a bandwidth of 38 nm defining the FUV band. A useful exposure of 11042.765 seconds was obtained from 15 orbits of the satellite. These observations covered a large field of view of $28^{\prime}$ (or $\sim 0.5 ^{\circ} $) with an angular resolution of $\sim 1.4''$. The ID of the observation is A05$\_$167T01$\_$9000004940, and the PI of the observations is H.K. Jassal.

\section{UVIT reduction pipeline}
 Level 1 (L1) data from the observation were obtained from the Indian Space Science Data Center (ISSDC) archive. \footnote{ISSDC Data archive: \url{https://astrobrowse.issdc.gov.in/astro_archive/archive/Home.jsp}} The L1 data contain information about the orbits and raw photon events detected in the FUV channels. The data reduction was done via a pipeline with the help of CCDLAB \cite{2021JApA...42...30P}. The process included aligning of all the different orbit images and combining them into a single image, which reduces the further rotational and translational movements of the sources in the image. The optimization process reduced the point spread function of all the bright sources, and the World Coordinate System solution was applied to obtain the coordinate grid of right ascension \& declination of the image. After finalizing the science product, we obtained the final FUV image of the field, which was ready for analysis.
 
\section{Data and image analysis}
\begin{figure*}[!htb]
 \vspace{20pt}
 \sidecaption
      \centering
      \includegraphics[width=0.70 \textwidth]{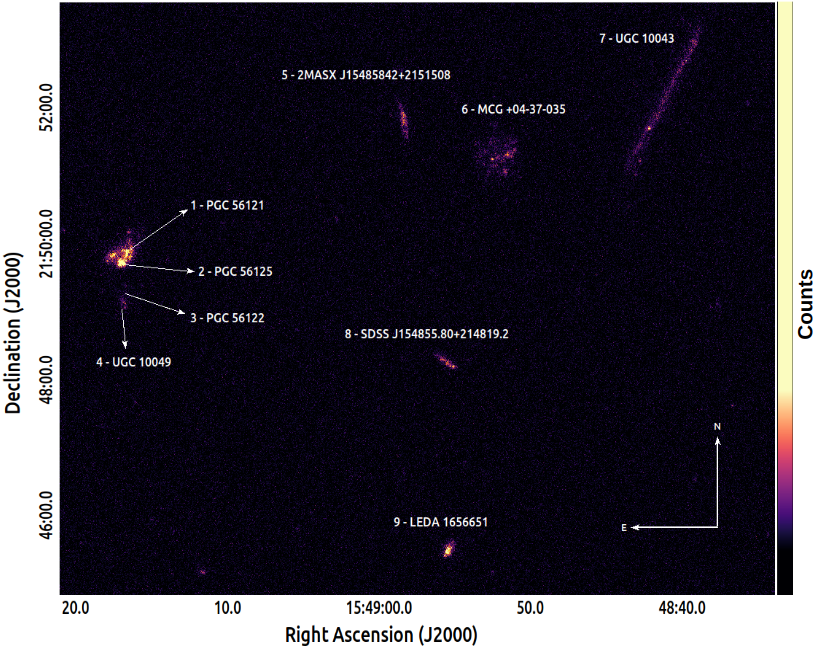}
      \caption{Far-UV images of galaxies in the field of view of the UVIT telescope while observing the HCG~77 region.}
      \label{objects}
  \end{figure*}

The FUV images of all the bright galaxies in the field of HCG~77 surveyed using the UVIT are shown in Figure \ref{objects}. Table \ref{tab:prop} lists basic information on the galaxies. In addition to our UVIT/\textit{AstroSat} observations, we utilized archival data from the SIMBAD Astronomical Database (\citet{Wenger_2000}), the NASA/IPAC Extragalactic Database (NED; \citet{1991ASSL..171...89H}), and the Sloan Digital Sky Survey (SDSS; \citet{Ahumada_2020}; \citet{2020AJ....160..120J}; \citet{2017ApJS..233...25A}). 
     \begin{table*}[!htb]
   \vspace{10pt}
\centering
\caption{ Properties of the galaxies observed in the UVIT field.}
\begin{tabular}{cp{3.5cm}p{2.5cm}cp{1.5cm}p{1.5cm}p{1.2cm}p{2.7cm}}
\hline
  No. & Name & RA(2000) \& Dec(2000) & Redshift & Distance (in Mpc) & Magnitude & Size (in kpc) & References \\
\hline
\hline
 1 & PGC~56121 & $15^{h}49^{m}16.65^{s}$, $+21^{d}49^{m}53.1^{s}$ & 0.006969 & $ 29.80$  & R = 15.02 & 1.30 & SDSS Data Release 13\\
 \hline
 2 & PGC~56125 & $15^{h}49^{m}17.10^{s}$, $+21^{d}49^{m}43.3^{s}$ & 0.007505 & $32.15$  & R = 16.58 & 0.90 & 1\\
\hline
3 & PGC 56122 & $15^{h}49^{m}16.80^{s}$, $+21^{d}49^{m}24.38^{s}$ & 0.035658 & 152.82  & R = 14.64 & 9.85 & 1\\
\hline
4 & UGC 10049 & $15^{h}49^{m}16.89^{s}$, $+21^{d}49^{m}09.87^{s}$ & 0.035051 & $148.92$ & R = 14.53 & 10.49 & 2\\
\hline
5 & 2MASX J15485842+2151508 & $15^{h}48^{m}58.43^{s}$, $+21^{d}51^{m}50.68^{s}$ & 0.024798 & $105.60$  & B = 16.62 R = 14.64 & 15.13 & SDSS Data Release 13\\
\hline
6 & MCG +04-37-035 & $15^{h}48^{m}52.19^{s}$, $+21^{d}51^{m}18.27^{s}$ & 0.007424 & $31.82$  & B = 16.21 R = 15.36 & 6.35 & SDSS Data Release 13 \\
\hline
7 & UGC 10043 & $15^{h}48^{m}41.12^{s}$, $+21^{d}52^{m}09.79^{s}$ & 0.007208 & $30.90$  & B = 14.8 & 21.36 & 3\\
\hline
8 & SDSS J154855.80+214819.2 & $15^{h}48^{m}55.87^{s}$, $+21^{d}48^{m}19.01^{s}$ & 0.024689 & $105.15$  & R=17.3 & 11.96 & SDSS Data Release 13\\
\hline
9 & LEDA 1656651 & $15^{h}48^{m}55.38^{s}$, $+21^{d}45^{m}34.64^{s}$ & 0.006693 & $28.68$ & R = 16.54 & 2.28 & SDSS Data Release 13 \\
\hline
\end{tabular}

\label{tab:prop}
\tablefoot{Some of the general characteristics listed here are available on SIMBAD (\url{https://simbad.u-strasbg.fr/simbad/sim-fbasic}) and NED-NASA/IPAC Extragalactic Database (\url{https://ned.ipac.caltech.edu/}). We used Lambda CDM cosmological parameters where the Hubble constant used for distance estimation from redshift is $70 \ km/s/Mpc$. For galaxies with regular morphology (e.g., spirals or ellipticals), sizes are measured along the semi-major axis. For galaxies with disturbed or irregular morphology, the sizes correspond to the maximum elongation visible in the image.}
\tablebib{(1)~\citet{1992ApJ...399..353H}; (2) \citet{1991rc3..book.....D}; (3) \citet{2005ApJS..160..149S}}
\end{table*}

In this Table, the first two objects listed are a possible interacting pair of galaxies, i.e., PGC~56121 and PGC~56125.  These two galaxies and those marked 3 and 4 are visible as two separate objects in the optical and 21 cm. \footnote{\url{https://www.nrao.edu/archives/items/show/33570}} The object at no. 5 is an isolated galaxy. Objects listed at 6 and 7 are observed to be physically connected by hydrogen gas (H~I; 21 cm), as initially presented by \citet{2009RMxAC..35..201A} and further characterized by \citet{Bahr_2025}. MCG +04-37-035 is a distorted galaxy interacting with UGC 10043, which is an edge-on barred spiral (type Sbc) galaxy in Figure \ref{objects}. Object 8 is an isolated galaxy in the frame, and the last object, LEDA 1656651, is a galaxy in a pair.

The point spread function value was determined by analyzing a bright point source ( RA: 15:50:04, DEC: +21:48:18) in the field of view image and measuring the intensity profile along the radial direction. We noted the counts from the pixel table from the center of the source to the endpoints in both directions, which are diametrically opposite. A Gaussian fit was plotted from the datasets, which helped in obtaining the standard deviation. The deviation value we obtained was 1.34 ($\sigma = 1.34 $) and the full width half maximum value was 3.15 ($FWHM = 2.354 \cdot \sigma = 3.15$). Since each value here denotes the pixel count, using the conversion factor for UVIT ($1 \ pixel = 0.4168 \ arcsecond$), we therefore obtained a final point spread function value of approximately 1.30 arcseconds for our FUV image, which is consistent with the expected resolution for the BaF\textsubscript{2} (F154W) filter. This confirmed that the image is well-calibrated and suitable for further analysis.

\begin{figure*}
 \begin{subfigure}[b]{0.48\textwidth}
    \centering
    \includegraphics[width=\linewidth]{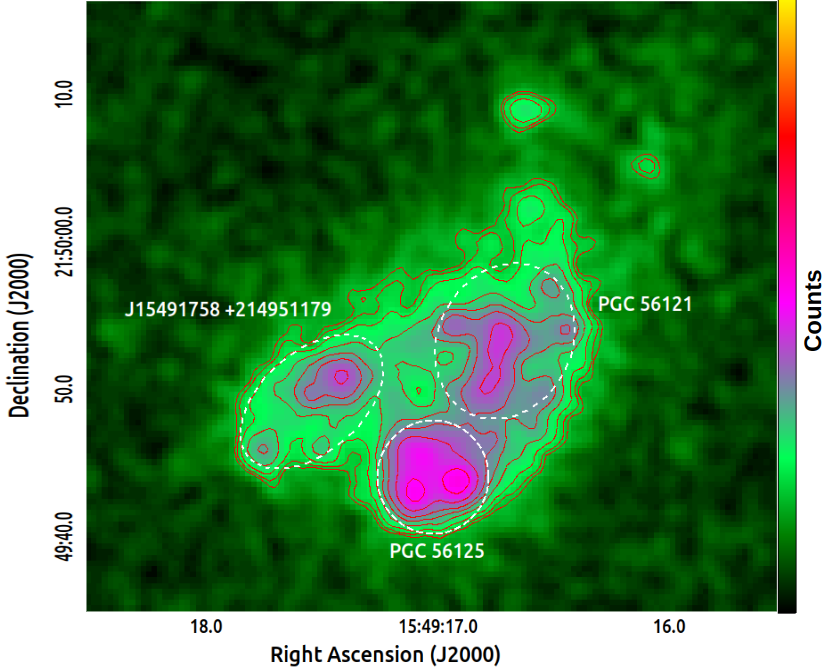}
    \caption*{}
   
\end{subfigure}
\begin{subfigure}[b]{0.48\textwidth}
     \centering
    \includegraphics[width=\linewidth]{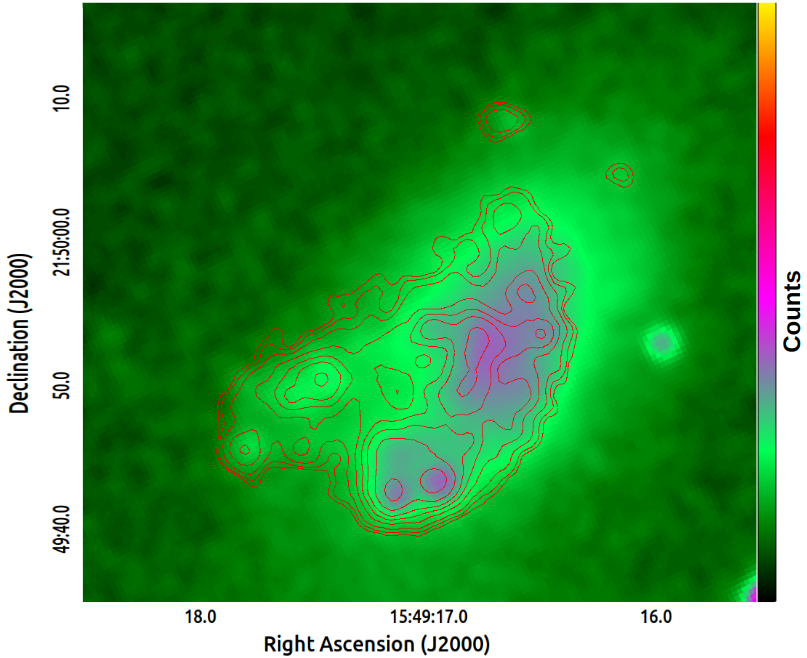}
    \caption*{}
    
\end{subfigure}
\caption{Surface brightness contour map of three sources : PGC~56121, PGC~56125, and J15491758 +214951179. \textit{Left}: The contour levels plotted are $4\sigma$, $5\sigma$, $7\sigma$, $11\sigma$, $16\sigma$, $22\sigma$, $34\sigma$, and $55\sigma$ above the background level on far-UV image. Regions highlighted in pink represent areas of higher flux density and are likely associated with active star-forming regions, while the green shaded areas denote more diffuse emission, possibly due to gas with lower luminosity. Objects analyzed in this study are outlined by white dashed regions. \textit{Right}: Contours derived from the FUV data overlaid on the SDSS r-band image for a comparison and to identify star-forming and brighter regions.}
\label{uvitHCG~77contour}
\end{figure*}

We obtained a well-resolved image of HCG~77 using UVIT on board {\it AstroSat}, which formed the basis of our analysis. To analyze the flux distribution and identify structural features in the interacting galaxy system HCG~77, we generated surface brightness contour maps using \textit{SAOImage DS9} (\citet{2003ASPC..295..489J}), shown in Figure \ref{uvitHCG~77contour}. We estimated the mean background level estimated to be 0.44, with a deviation of 0.35. These contours trace regions of equal flux, allowing us to distinguish between dense star-forming zones and more diffuse extended emission. Higher contour levels correspond to regions of greater luminosity, which are likely indicative of ongoing star formation, while lower levels highlight fainter and possibly gas-dominated areas. This visualization aids in understanding the spatial distribution of activity across the system. The figure also marks the three primary regions examined in detail: the galaxies PGC~56121 and PGC~56125 along with J15491758 +214951179. The source J15491758 +214951179 exhibits a tidal tail-like feature, which appears to be emerging from the parent galaxy, suggesting it is a coherent part of the tidal structure originating from PGC~56121. These sources constitute the main focus of this study.

\subsection {Photometry}
After obtaining the complete image, we performed aperture photometry using Source Extractor (SEXTRACTOR) software \citet{1996A&AS..117..393B}. In this process, we detected and extracted the sources that were responsible for emitting flux during galaxy interactions. We extracted the sources from a subsection of the entire FUV image centered on the HCG~77 group.
The minimum number of pixels to be considered a detection ($DETECT\_MINAREA$) was set to seven, which is approximately equal to the area of the circle with a diameter 3 pixels (which is the value of the point spread function we obtained). We kept the value $DETECT\_THRESH$ $3  \sigma$ so that SEXTRACTOR detects objects with flux values of $3  \sigma$ or above, where $\sigma$ is the average background noise. Although lower thresholds capture extended sources, SEXTRACTOR would also take into account noises, and a higher threshold may reduce false detections, SEXTRACTOR but would also miss faint sources, so the value $ 3  \sigma$ is good to start with. 
For the background parameters, we took $BACK\_TYPE = AUTO$ and $BACK\_SIZE = 64\ pixels$. 
The value of the zero-point magnitude ($MAG\_ZEROPOINT$) we used was 17.78 for FUV $BaF_{2} \ (F154W)$, which was taken from \citet{Tandon_2020}, where they provided additional calibration of UVIT on board {\it AstroSat}.
For the apertures, we used ellipses that have appropriate values of minimum radius, the Kron factor, and the Petrosian factor. Several key parameters used in our analysis were adopted from studies based on FUV observations by \citet{Samantaray_2024} and \citet{Mahajan_2022}. The parameters used for source extraction are listed in Table \ref{SExtractorTable}. 
\begin{table}[!h]
\centering
	\caption{SEXTRACTOR parameters used for the analysis of the UVIT FUV images.}
	{
	\begin{tabular}{lc} 
		\hline
	Main Parameters & Values \\
		\hline	
            \hline
	DETECT\_MINAREA & 7   \\
	DETECT\_THRESH ($\sigma$) & 3 \\	
   FILTER\_NAME  & gauss\_3.0\_7x7.conv   \\ 
   DEBLEND\_NTHRESH  & 32.0  \\ 
   DEBLEND\_MINCONT  & 0.005 \\ 
   CLEAN\_PARAM & 1.0 \\ 
  PHOT\_APERTURES  & 4  \\ 
PHOT\_AUTOPARAMS &  0.5, 1.0  \\
PHOT\_PETROPARAMS &  0.5, 1.0 \\
MAG\_ZEROPOINT & 17.78\\
GAIN & 1.0\\
PIXEL\_SCALE &  0.4168 \\
STARNNW\_NAME &  default.nnw \\
BACK\_TYPE &  AUTO \\
BACK\_SIZE & 64 \\

	\hline  
\end{tabular}}
\label{SExtractorTable}

\end{table}

 Source Extractor was used to identify sources in the image and to obtain initial estimates of the flux contributing sources through elliptical apertures.  While SEXTRACTOR is most effective for detecting point sources and compact galaxies in wide fields, studying resolved structures within a galaxy and investigating individual star-forming regions required us to manually define appropriate regions of interest by looking at the images and performing photometric measurements directly. SEXTRACTOR served primarily as an indicative tool to locate possible sources, while detailed analysis of individual regions was carried out manually.
\subsection{SFR determination}
\begin{figure}[!htb]
      \centering
      \includegraphics[width=1.0 \columnwidth]{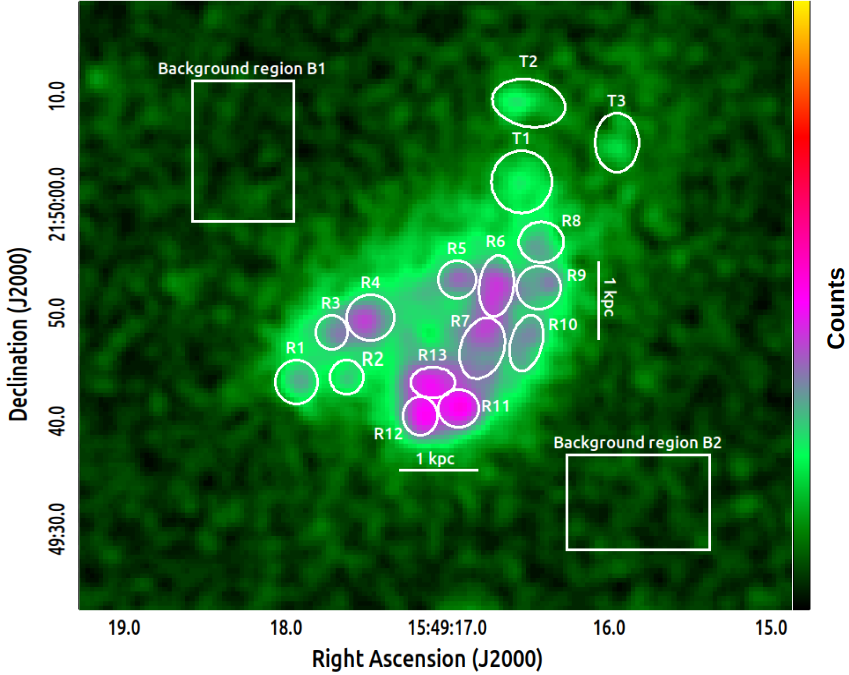}
      \caption{Star-forming regions identified within the diffuse gas of the system. Regions R1–R13 mark the star-forming sites detected through photometry. Background regions B1 and B2 were selected to estimate the background counts. Regions T1-T3 represent a gas trail that is also associated with star formation.}
      \label{regions}
  \end{figure}
Figure \ref{regions} depicts the star-forming regions identified with the help of \textit{Source Extractor} along with those defined manually. The background regions provide the background value, which can be used to determine the correct fluxes of star-forming regions.
After identifying the star-forming regions, we determined the SFRs in these regions. 
The FUV-based SFR was calculated using the relation provided by \citet{iglesias2006star}:
\begin{equation}
     \text{log}(\text{SFR}_{\text{FUV}}/\text{M}_{\odot} \text{yr}^{-1}) = \text{log}(\text{L}_{\text{FUV}}/\text{L}_{\odot}) - 9.51
     \label{SFRrate}
\end{equation}
Luminosities of the regions were calculated by the counts value and then converted into the flux as prescribed in the AstroSat instrument calibration.\footnote{Conversion factors for different filters are available at \url{https://www.iiap.res.in/projects/uvit/instrument/filters/}} Since our analysis was based on the FUV BaF\textsubscript{2} (F154W) filter image, we adopted the values of the unit conversion factor, mean wavelength ($\lambda_{\text{mean}}$), and bandwidth ($\Delta\lambda$) from the provided table. The extinction $A_{FUV}$ was also taken into account, which is given by $A_{FUV}=8.06 \times E(B-V)$ following \citet{bianchi2011galex}. The reddening value for HCG~77 was obtained from the IRSA Dust Extinction Service, \footnote{ \url{https://irsa.ipac.caltech.edu/applications/DUST/}} which provides values based on \citet{schlafly2011measuring}. We adopted their result, yielding a reddening of $E(B-V) \approx 0.05$, and consequently obtained an extinction value $A_{FUV} \approx 0.396$. 
 The details of the area of the star-forming regions, their luminosities, and SFRs are given in Table \ref{tab:region_data}. 
 \begin{table}[!h]
\centering
\footnotesize
\caption{Area, luminosity, and SFR of the identified star-forming regions.}
\label{tab:region_data}
\begin{tabular}{cccc}
\hline
Region & Area (kpc$^2$) & Luminosity ($10^6 L_{\odot}$) & SFR ($10^{-3} M_{\odot}$/yr) \\
\hline
\hline
R1 & 0.272 & 1.313 & 0.406 \\
R2 & 0.154 & 0.740 & 0.229 \\
R3 & 0.147 & 1.244 & 0.385 \\
R4 & 0.309 & 3.431 & 1.061 \\
R5 & 0.198 & 1.946 & 0.602 \\
R6 & 0.279 & 3.994 & 1.235\\
R7 & 0.368 & 4.444 & 1.374 \\
R8 & 0.250 & 1.525 &  0.472 \\
R9 & 0.265 & 2.183 & 0.675 \\
R10 & 0.254 & 2.119 & 0.655 \\
R11 & 0.217 & 5.919 & 1.830 \\
R12 & 0.191 & 4.851 & 1.499 \\
R13 & 0.177 & 4.320 & 1.336 \\
T1 & 0.364 & 1.145 & 0.354 \\
T2 & 0.313 & 0.781 & 0.241 \\
T3 & 0.235 & 0.395 & 0.121 \\
\hline
\end{tabular}

\end{table}

\section {Spectral energy distribution}
We modeled the SED of the two galaxies PGC~56125 and PGC~56121 and the source J15491758 +214951179 using ultraviolet data from our analysis complemented by archival optical and infrared data. The resulting SED modeling provides key physical parameters of galaxies, including stellar mass, luminosity, SFR, metallicity, and the chemical composition of their surrounding environment. 

\subsection{Dataset}
We obtained the optical and infrared dataset from the \textit{SDSS} and \textit{Vizier} photometry database via \textit{SIMBAD}. 
For the optical band, we used SDSS data containing five filters: $u,~g,~r,~i$, and $z$. The information of the fluxes and errors was obtained from the FITS image of five SDSS filters.
For the infrared band, we used WISE data containing $W1, W2, W3$, and $W4$ filters, and the information about their exposure times, fluxes, errors, and wavelengths was obtained from the Vizier catalog. \footnote{\url{http://vizier.cds.unistra.fr/vizier/sed/?submitSimbad=Photometry}} \\

We modeled the SEDs of three sources: one with a semimajor axis of $\sim 6''$ and a  semiminor of $\sim 5''$ centered at $237.3199421^\circ$, $21.8311516^\circ$ for PGC~56121; another of radius $\sim4.2''$ centered at $237.3213249^\circ$, $21.8288304^\circ$ for PGC~56125; and the last with a semimajor axis of $\sim6.3''$ and a semiminor axis of $\sim3.7''$ centered at $237.3239517^\circ$, $21.8304228^\circ$ for J15491758 +214951179. We gathered the properties and dynamics of the system. The measured values of the fluxes and errors in the different filters are given in Table \ref{tab:photometry}.\\
\begin{table*}[!htb]
\caption{Photometric observations used in the SED fitting for PGC~56121, PGC~56125, and J15491758 +214951179}
\label{tab:photometry}
\centering
\begin{tabular}{lccccccc}
\hline
\multicolumn{2}{c}{} & \multicolumn{2}{c}{PGC~56121} & \multicolumn{2}{c}{PGC~56125} & \multicolumn{2}{c}{J15491758 +214951179}\\
\hline 
\hline
Filter & Wavelength ($\mu$m) & Flux (Jy) & Error (Jy) & Flux (Jy) & Error (Jy) & Flux(Jy) & Error(Jy) \\
\hline
{\it AstroSat} F154W & 1.54E-1 & 7.33E-05 & 2.34E-06 & 6.79E-05 &  2.28E-06 & 2.34E-05 & 1.82E-06\\
SDSS $u$       & 3.52E-1 & 2.78E-04 &  3.74E-05  & 2.16E-04 &  3.13E-05 & 9.06E-05 & 2.25E-05\\
SDSS $g$       & 4.82E-1 & 6.55E-04  & 6.72E-05  & 3.34E-04  & 4.59E-05 & 1.44E-04 & 3.78E-05 \\
SDSS $r$       & 6.25E-1 & 9.14E-04  & 8.53E-05 & 3.93E-04  & 5.40E-05 & 1.71E-04 & 4.71E-05\\
SDSS $i$       & 7.63E-1 & 1.06E-03 &  9.78E-05  & 3.66E-04  & 5.75E-05 & 1.87E-04 & 5.45E-05\\
SDSS $z$       & 9.02E-1 & 1.27E-03 &  1.04E-04  & 3.98E-04  & 6.18E-05 & 1.96E-04 & 5.93E-05\\
WISE W1        & 3.35E+0 & 3.45E-04  & 8.00E-05  & 4.40E-04 &  1.30E-05 & - & -\\
WISE W2        & 4.60E+0 & 2.57E-04  & 2.80E-05  & 3.00E-04 &  1.30E-05 & - & -\\
WISE W3        & 1.16E+1 & 4.93E-04  & 8.60E-05   & 9.92E-04 &  9.20E-05 & - & -\\
WISE W4        & 2.21E+1 & 3.70E-03  & 7.80E-04  & 4.80E-03 &  6.80E-04 & - & - \\
\hline
\end{tabular}
\tablefoot{This table provides fluxes and errors for both galaxies across different filters used in the SED modeling. The fluxes and errors of the {\it AstroSat} F154W filter and SDSS filters were measured from the FITS image; the rest were obtained via the VizieR survey catalogs.}
\end{table*}

Photometric measurements were carried out on the FUV image obtained with UVIT using a fixed aperture size  centered on the galaxy. The same aperture size and position were adopted for all other bands to ensure that the derived fluxes correspond to the same projected region on the sky. Multiwavelength photometric data from surveys such as SDSS and WISE were obtained via the VizieR photometry services using the same aperture radius. The selected aperture size was sufficiently chosen to minimize potential flux losses or color biases.

For the calculation of flux for the FUV F154W filter, we used the expression\begin{equation}
    F_{\nu}=\frac{\lambda^{2}}{c}\times CPS \times UC
\end{equation}
where $\lambda$ is the mean wavelength of the filter. Here, it is 1541 \AA. The term ``CPS'' is the counts per second measured from the SAOImage DS9 region statistics, which takes into account background counts and errors, and ``UC'' is the unit conversion factor in ergs per second per centimeter square per angstrom from AstroSat instrument calibration, \footnote{\url{https://www.iiap.res.in/projects/uvit/instrument/filters/}} and finally we calculated $F_{\nu}$ and its error in Janskys. For the calculation of flux for the SDSS filters, the \textit{BUNIT} was given in ``nanomaggy," which describes the physical counts value per pixel. The total number of counts gives the total nanomaggy, which can be converted into flux using the conversion $1\ \mathrm{nanomaggy}=3.631\times10^{-6} \mathrm{Jy}$ (\citet{2022ApJS..259...35A}).

\subsection{Modeling the SED}
We modeled the SED of the galaxies using the Multi-wavelength Analysis of Galaxy Physical Properties (MAGPHYS) package (\citet{da_Cunha_2008}). It allowed us to interpret observations for a wide range of wavelengths ($912\AA \leq \lambda \leq 1 mm $) along with the model libraries -- optical and infrared. The optical model accounts for stellar emission from galaxies along with dust attenuation as described in \citet{Charlot_2000}. The infrared model libraries account purely for dust emission. The interstellar medium (ISM) of the galaxies is described by two components: the ambient (diffuse) ISM and the star-forming region (\citet{da_Cunha_2011}). The combined models and libraries give the full SED of our interacting system at wavelengths from the FUV to the mid-infrared.

\begin{figure}[!ht]
\vspace{20pt}
    \begin{subfigure}[b]{1.0 \linewidth}
        \centering
        \includegraphics[width=1.0 \linewidth]{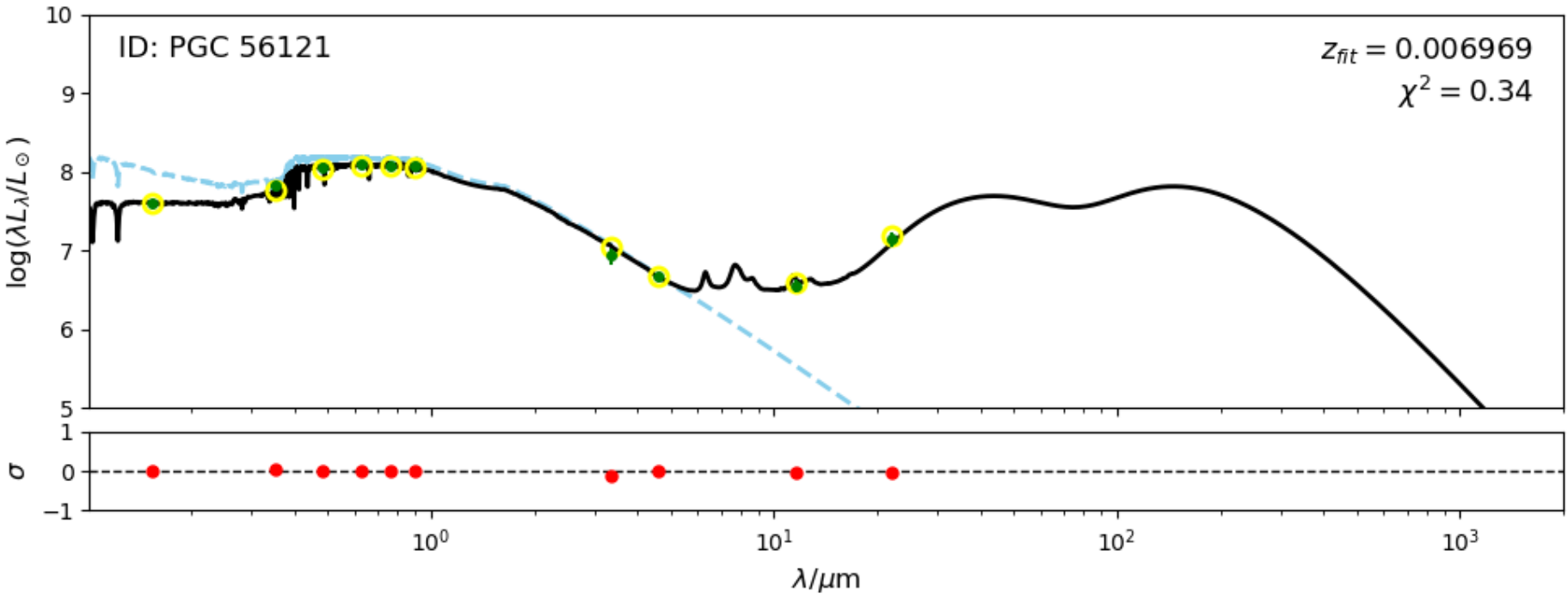}
        \caption*{}
    \end{subfigure}
    \hfill
    \begin{subfigure}[b]{1.0\linewidth}
        \centering
        \includegraphics[width=1.0 \linewidth]{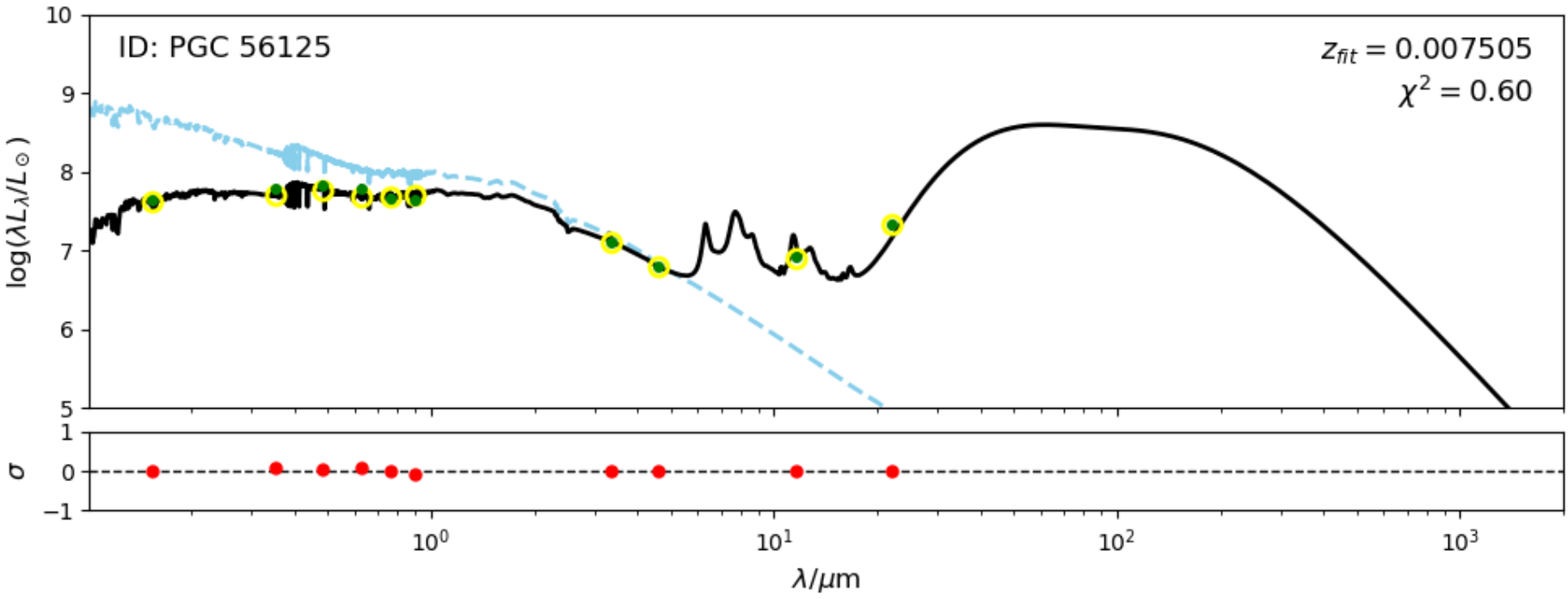}
        \caption*{}
    \end{subfigure}
    \hfill
    \begin{subfigure}[b]{1.0 \linewidth}
        \centering
        \includegraphics[width=1.0 \linewidth]{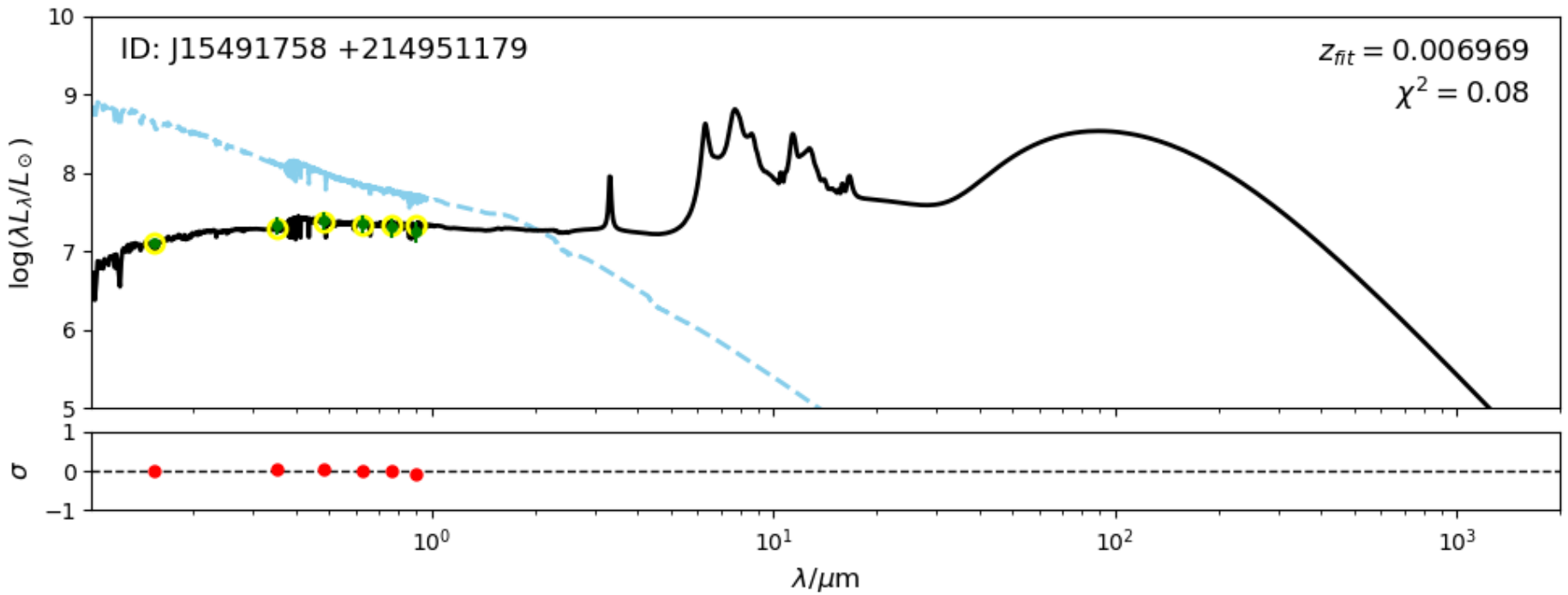}
        \caption*{}
    \end{subfigure}
\caption{Integrated SED of the system of two galaxies, PGC~56121 (\textit{top}) and PGC~56125 (\textit{middle}), and of J15491758 +214951179 (\textit{bottom}) from FUV to mid-infrared region. In all the figures,the \textit{top} panel shows the photometric points in green dots from Table \ref{tab:photometry}, and yellow circles are model photometric points at the same wavelength. The black line shows the best-fit model curve across all wavelengths in this regime, and the dashed blue line represents the unattenuated stellar emission. Here, $\chi^{2}$ depicts the total chi-square value of best-fit model and $z_{fit}$ is the redshift of the model used. The bottom panel shows the residuals of the photometric points in red dots along the black dashed line at y=0.}
\label{sed}
\end{figure}

We obtained SEDs from ten photometric data points and derived the best-fit models that give the best-fit values of physical parameters -- stellar mass ($M_{*}$), dust mass ($M_{dust}$), SFR, dust luminosity ($L_{dust}$), and among others. The model libraries were generated by MAGPHYS at redshifts 0.0070 and 0.0075 for analysis, which resulted in an integrated SED. The SEDs are shown in Figure \ref{sed} with a total chi-square value of 0.34 and  0.60 and at a redshift of $z_{fit}= 0.006969$ and 0.007505 for PGC~56121 and PGC~56125, respectively. For the analysis of J15491758 +214951179, we took $z_{fit}= 0.006969$ as inferred from the H~I velocity field in \citet{2009RMxAC..35..201A}. In Figure \ref{sed}, the chi-square value corresponds mainly to the optical chi-square, as the majority of our photometric points lie at shorter wavelengths, resulting in the fit being constrained predominantly by the optical data.

\begin{figure*}[!htb]
    \centering

    \begin{subfigure}[b]{0.48\textwidth}
       \centering
        \includegraphics[width=\linewidth]{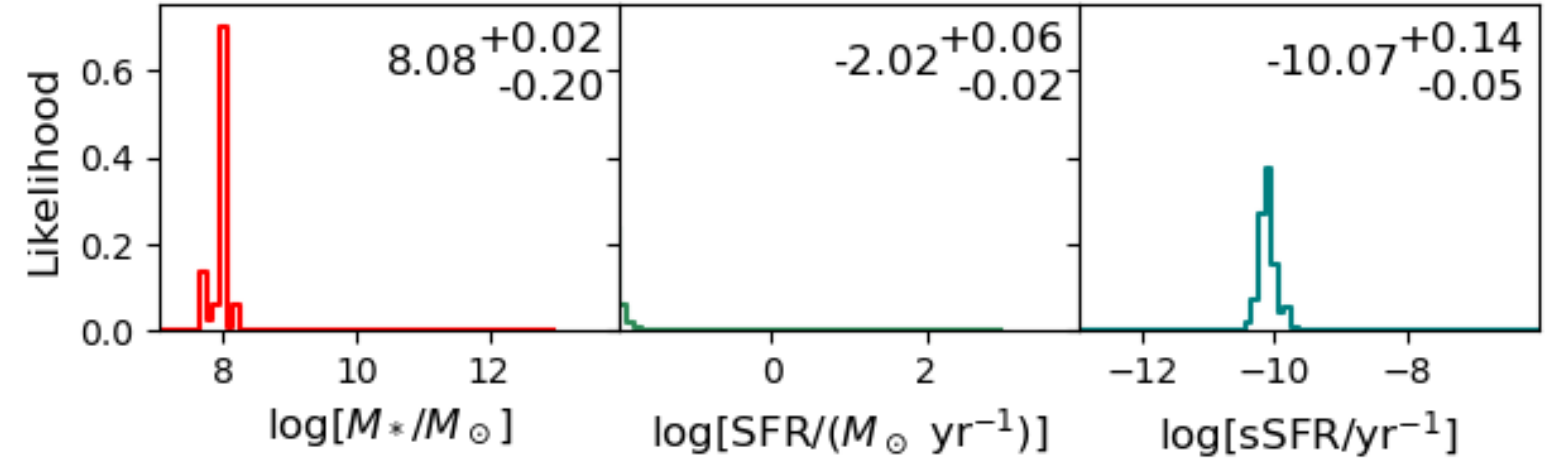}
        \caption*{}
    \end{subfigure}
    \hfill
    \begin{subfigure}[b]{0.48\textwidth}

        \centering
        \includegraphics[width=\linewidth]{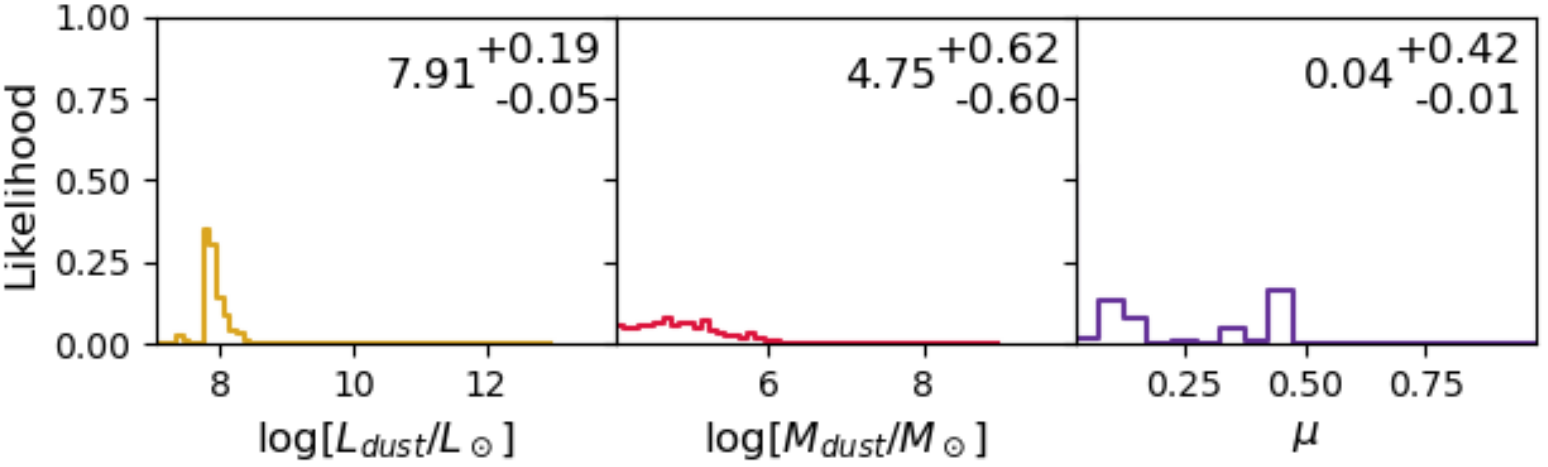}
        \caption*{}
    \end{subfigure}

    \vspace{0.1cm} 
    \begin{subfigure}[b]{0.48\textwidth}

        \centering
        \includegraphics[width=\linewidth]{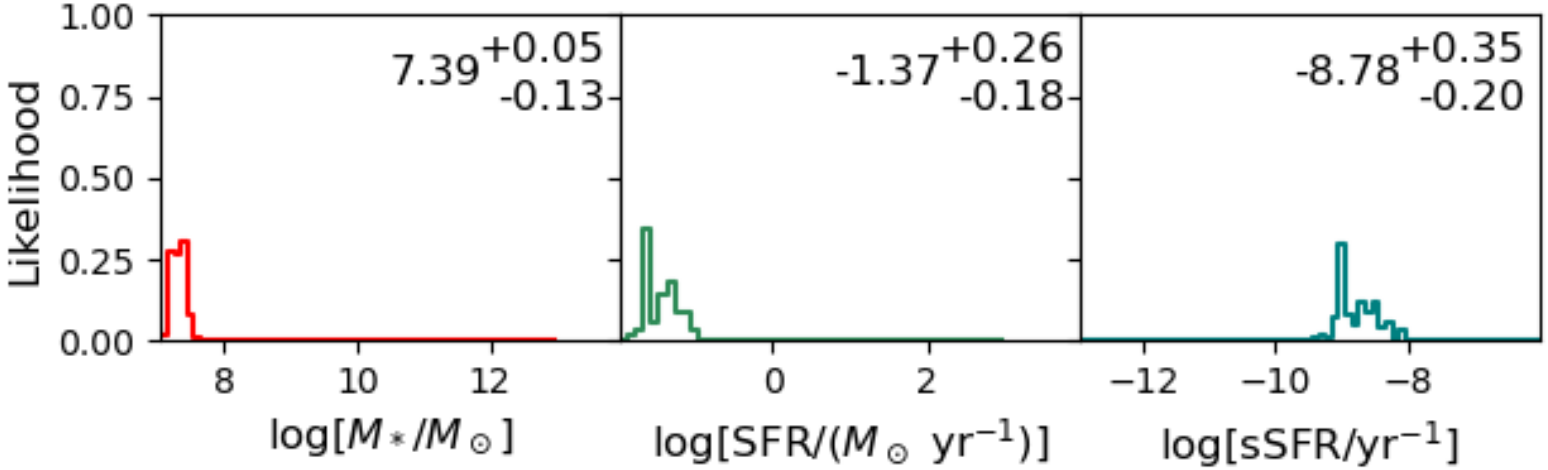}
        \caption*{}
    \end{subfigure}
    \hfill
    \begin{subfigure}[b]{0.48\textwidth}

        \centering
        \includegraphics[width=\linewidth]{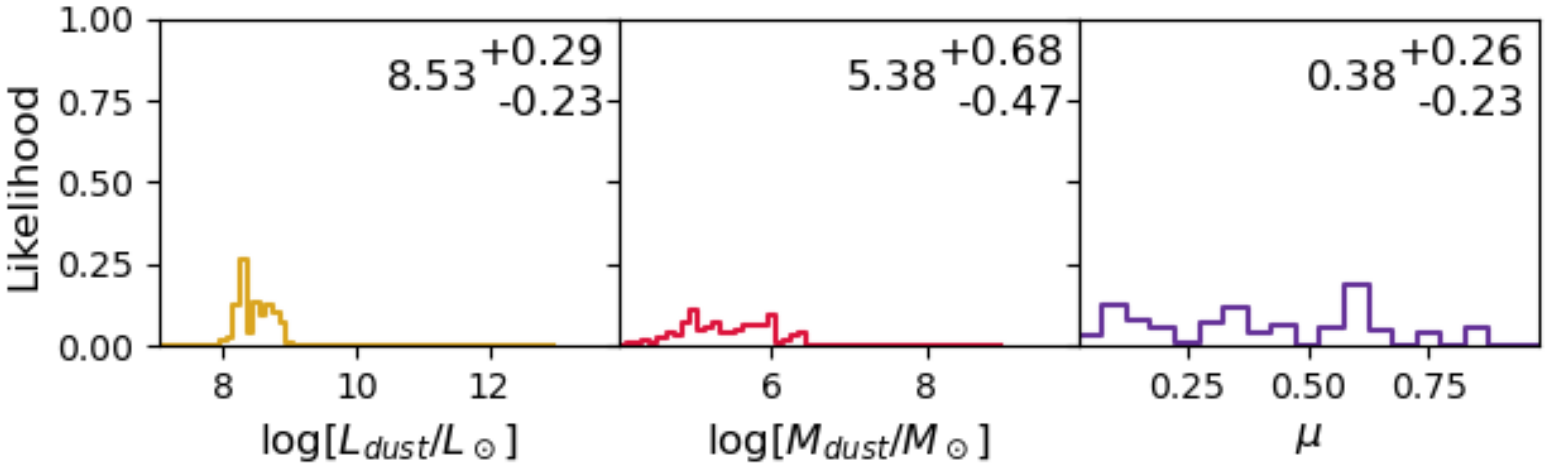}
        \caption*{}
    \end{subfigure}

    \vspace{0.1cm} 
    \begin{subfigure}[b]{0.48\textwidth}

        \centering
        \includegraphics[width=\linewidth]{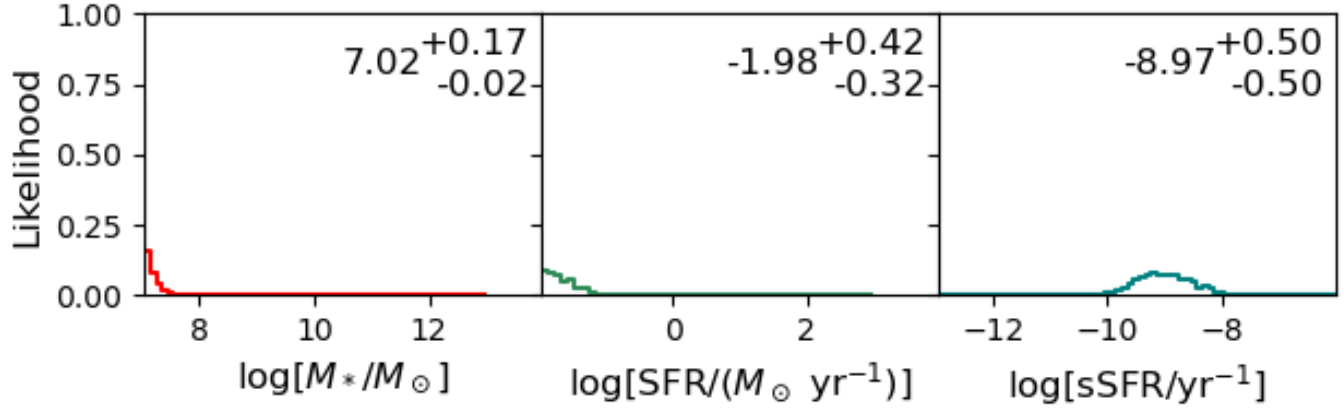}
        \caption*{}
    \end{subfigure}
    \hfill
    \begin{subfigure}[b]{0.48\textwidth}

        \centering
        \includegraphics[width=\linewidth]{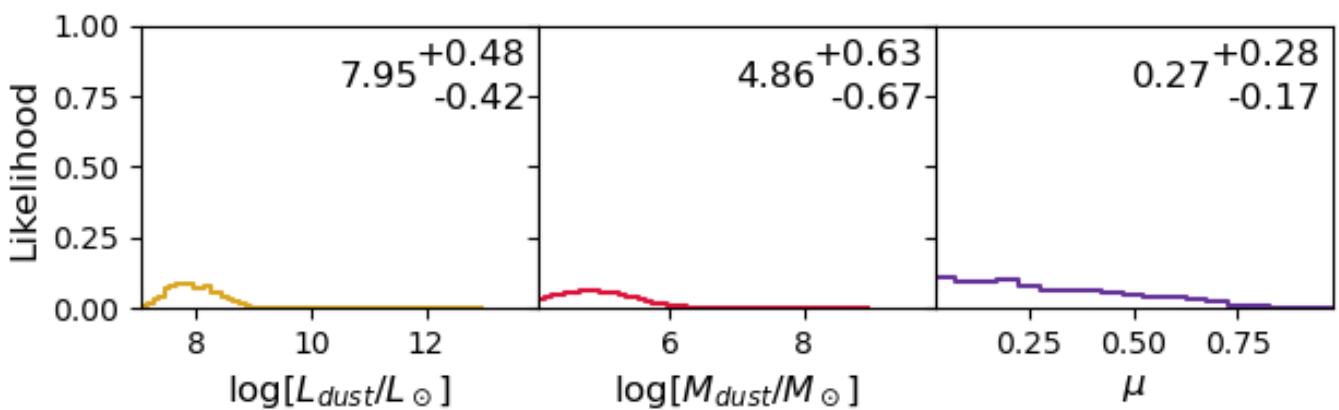}
        \caption*{}
    \end{subfigure}

    \caption{Relative probability distributions of various physical parameters derived from SED fitting. The top row shows the likelihood curves of physical parameters for galaxy PGC~56121, and the middle and bottom rows show those for PGC~56125 and J15491758 +214951179, respectively. The best-fit values and their associated $16^{\text{th}}$ and $84^{\text{th}}$ percentile uncertainties are indicated.}
    \label{likelihood}
\end{figure*}

The SED modeling provides values of parameters that are important tools for understanding the dynamics and properties of the system. Figure \ref{likelihood} shows the distribution and median values of various parameters along with an uncertainty of $\pm1\sigma$.

The $\mu$ parameter provides information about the fraction of dust attenuation in the ISM, with its two main components being the ambient (diffuse) ISM and the star-forming clouds. It is given by
\begin{equation}
    \mu=\frac{\hat\tau_{V}^{ISM}}{\hat\tau_{V}^{ISM}+\hat\tau_{V}^{BC}}
\end{equation}
Here, $\hat\tau_{V}^{BC}$ is the effective V-band absorption optical depth of the dust in stellar birth clouds, and $\hat\tau_{V}^{ISM}$ is the same but in the ambient ISM (\citet{da_Cunha_2008}).
This parameter indicates the percentage of dust received from the ambient ISM. 
For PGC~56121, $\mu$ is 0.460; and for PGC~56125, it is 0.866.  
Models in the system are described with an SFR  decaying exponentially as
\begin{equation}
    \psi(t)\propto exp(-\gamma t)
\end{equation}
where $\gamma$ is the star formation timescale parameter (in gigayear inverse) describing the star formation history(\citet{da_Cunha_2008}).
 We obtained the values of $\gamma$ to be $0.086\ \mathrm{Gyr}^{-1}$ for PGC~56121 and $0.842\ \mathrm{Gyr}^{-1}$ for PGC~56125. The onset of star formation was estimated to have occurred approximately 4.5 billion years ago for PGC~56121 and around 6.7 billion years ago for PGC~56125. The metallicity of PGC~56121 was found to be ($Z/Z_{\odot}$) = 0.022, while that of PGC~56125 was ($Z/Z_{\odot}$) = 1.906. Other important physical parameters derived from the SED and likelihood distribution function are given in Table \ref{physicalparameters}.

We also performed a g–r color–magnitude analysis of the galaxy PGC~56121 and J15491758 +214951179. The AB magnitude of the SDSS filter used is given by the relation
\begin{equation}
    m=22.5-2.5log_{10}[f]
\end{equation}
where ``f'' is the flux in ``nanomaggy''(\citet{2022ApJS..259...35A}). Using the \textit{g} and \textit{r} filters, we obtained the g-r color of PGC~56121 to be $\sim0.355$. 
For J15491758 +214951179, we obtained a g-r color of $\sim 0.041$. The g-r color offset between the PGC~56121 and J15491758 +214951179 is $\sim0.314$.

\section{Discussion}

Based on existing observational data, we interpret HCG~77 as primarily consisting of a possibly interacting pair, PGC~56121 and PGC~56125, rather than the traditionally assumed four-member system. Although UGC~10049 and PGC~56122 have often been associated with HCG~77 in earlier literature, their redshifts, as reported in archival data and also summarized in Table \ref{tab:prop}, differ significantly from those of PGC~56121 and PGC~56125. This suggests that they are not physically bound to the group. These findings and our analysis support a more restricted interpretation of HCG~77 as a system dominated by PGC~56121, PGC~56125, and a candidate TDG J15491758 +214951179.

From the constructed SED of the interacting system, we derived several key physical parameters. The estimated stellar masses, SFRs, and physical sizes of the galaxies PGC~56121 and PGC~56125 are broadly consistent with those reported by \citet{Amrutha_2024}. Based on these parameters, along with morphological features, we classify  PGC~56121 as an irregular barred magellanic-type galaxy, and this is supported by the presence of a visually identifiable bar-like structure. In contrast, PGC~56125 exhibits a compact morphology and active star formation, characteristics consistent with being a blue compact dwarf galaxy.

The dust attenuation parameter \( \mu \) of PGC~56121 is found to be 0.46, implying that about 46\% of the dust emission comes from the ambient ISM. This suggests a significant contribution from birth clouds and a relatively younger stellar population in PGC~56121 than in PGC~56125, whose \(\mu\) is 0.86.

For PGC~56121, the star formation timescale parameter \( \gamma = 0.086\,\text{Gyr}^{-1} \) indicates that the SFR is slowly declining over time, which is consistent with a system experiencing sustained star-forming activity. For PGC~56125, \(\gamma\) is 0.842 $Gyr^{-1}$, indicating that the star formation activity is declining and the galaxy may be heading toward quenching. The estimated metallicity of PGC~56121 is 0.022 times the solar value, pointing to a metal-poor environment with a high gas fraction. The recognition that such galaxies can simultaneously exhibit low metallicity and high SFRs originates from the studies of \citet{1972ApJ...173...25S}, as referenced in \citet{Kunth_2000}. The metallicity of PGC~56125 is 1.906 $Z_{\odot}$, indicating a metal-rich galaxy.

Located at the tip of the tidal tail of PGC~56121, J15491758 +214951179 exhibits properties indicative of a possible TDG. We obtained a g–r color of $\sim$ 0.355 for PGC~56121 and $\sim$ 0.041 for J15491758 +214951179, and the resulting color offset is $\sim$ 0.314. This significant difference suggests a younger stellar population in the tidal feature. Additionally, the stellar mass of J15491758 +214951179 is \( 6.02 \times 10^6\,M_\odot \), which is less than 10 \% of the stellar mass of the parent galaxy, and it is consistent with the expectations for tidal dwarfs candidates, as these findings align with the statistical criteria outlined in \citet{Kaviraj_2011} to identify TDGs. The findings also match the observations and results stated in \citet{Zaragoza_Cardiel_2024}. The most extended feature within this tidally influenced region measures roughly 0.9 kpc.

\begin{table*}[!htb]

\centering
\caption{Physical parameters obtained from the SED.}\label{physicalparameters} 
\begin{tabular}{lccc}
\hline
      Parameter    &PGC~56121&PGC~56125&J15491758 +214951179\\
\hline
\hline

Stellar mass $M_{*}$&$1.21\times10^8 M_{\odot}$&$2.63\times10^7 M_{\odot}$&$6.02\times10^6 M_{\odot}$
\\
\\
Dust mass $M_{dust}$ & $6.46\times10^5M_{\odot}$ & $1.21\times10^6M_{\odot}$& $6.28\times10^5 M_{\odot}$\\
\\
Dust luminosity $L_{dust}$& $1.26\times 10^8 L_{\odot}$&$6.66\times10^8L_{\odot}$& $6.83\times10^8L_{\odot}$\\
\\
Star formation rate SFR& $1.16\times10^{-2}M_{\odot}/yr$&$7.92\times10^{-2}M_{\odot}/yr$ & $5.20\times10^{-2}M_{\odot}/yr$\\
\\
Specific SFR&$9.16\times10^{-11}yr^{-1}$&$3.02\times10^{-9}yr^{-1}$ & $8.45\times10^{-9}yr^{-1}$\\
\hline
\end{tabular}
\end{table*}

The SED fitting for the TDG candidate was performed using only optical photometric data, as infrared measurements were not available for this region. The absence of mid- and far-infrared fluxes may lead to an underestimation of the total stellar mass and dust content. Consequently, the derived stellar mass should be considered as an indicative value rather than a precise measurement. However, even within this limitation, the estimated stellar mass ($M_{*} \sim 10^{6} M_{\odot}$) remains consistent with the mass range typically observed for TDGs. Moreover, even if the true stellar mass were higher by up to an order of magnitude, it would still lie within the broad range reported for TDG candidates in the literature, thus supporting the classification of this object as a TDG candidate.

The star-forming regions R8 to R10 in Figure \ref{regions} are relatively faint in comparison to R6 and R7, which trace the central barred part of the galaxy, and the photometric data available for them were insufficient to reliably construct their SEDs. Looking at their structural alignment, we interpret regions R8 to R10 as being part of the extended outer arm of PGC~56121. In addition, regions T1–T3 might correspond to gas streams or be a part of an outer arm of star-forming regions in PGC~56121 going in the northwest direction. However, more data, deeper multiwavelength imaging, or spectroscopy are needed to better understand the nature of these regions and to confirm their physical association with the parent galaxy.

Multiple indicators suggest that the five galaxies — UGC 10043, MCG+04-37-035, LEDA 1656651 (also referred as J1592443+215115), and the HCG~77 system (PGC~56121 and PGC~56125) are engaged in mutual interactions and may collectively constitute a galaxy group. Archival Very Large Array (VLA) radio data show clear evidence of UGC 10043, the most massive galaxy among them, drawing gas from MCG+04-37-035, which is consistent with tidal interactions (\citet{Matthews_2004}). Furthermore, velocity contour maps reveal an asymmetric distribution of neutral hydrogen around HCG~77 and a trail of HI material extending from LEDA 1656651 to UGC 10043, reinforcing the case for ongoing interactions (\citet{2009RMxAC..35..201A}).

In terms of projected separations, the distances among these objects lie between $\sim0.33 Mpc$ and $\sim3.5 Mpc$. These spatial proximities, along with the observed signs of interaction, also suggest that these galaxies may constitute a small dynamically connected group. The morphological distortions observed in the HCG~77 system, particularly in PGC~56121 and PGC~56125, may be attributed to past or ongoing gravitational interactions with these galaxies, which together comprise a larger-scale group. The SFRs on the order of $10^{-2} M_{\odot}\ yr^{-1}$
 observed in our dwarf galaxies, with stellar masses in the range $10^7-10^8 M_{\odot}$, are consistent with values typically found in star-forming dwarf systems. 
 
 The dwarfs in our study lie in proximity to a larger barred spiral galaxy, which suggests that tidal perturbations from the massive companion may be influencing their evolution. Previous observational work has shown that dwarfs located near massive spirals often display enhanced star formation, tidal disturbances, and morphological asymmetries consistent with interaction-driven activity (\citet{10.1093/mnras/stu1804}). The star formation we measured is therefore plausibly linked to external tidal effects rather than purely internal processes, highlighting the important role of galaxy environment in regulating the growth and star-forming properties of dwarfs. These observations suggest the presence of an evolving group structure influencing the dynamics and star formation activity.

\section{Summary}
We have presented a high resolution FUV image of galaxies in and around the group HCG~77, using UVIT on board {\it AstroSat}. While  HCG~77 has traditionally been considered a four-member group, existing redshift data, previous classifications, and the analysis presented in this study indicate that it is more accurately described as  a system consisting of PGC~56121, PGC~56125, and J15491758 +214951179 (TDG). The first two galaxies are at redshifts of 0.006969 and 0.007505, respectively, whereas the accurate redshift of the TDG is still unknown. We respectively classify PGC~56121 and PGC~56125 as an irregular barred magellanic galaxy and a blue compact dwarf based on their compact sizes, their low stellar masses, and the SFRs derived in this study. The classifications are also supported by the morphological appearance of the galaxies.

We report the presence of a possible TDG at the end of the tidal tail of the parent galaxy PGC~56121, identified as J15491758 +214951179. This source merits further study across multiple wavelengths, and accurate redshift measurement is required. The physical extent of these galaxies is approximately 1 kpc. Other galaxies within the field of view are significantly larger compared to PGC~56121 and PGC~56125. The edge-on barred spiral galaxy UGC 10043 is the largest in the field of view, with a diameter of 21.36 kpc.  Additional signs of interaction with nearby galaxies, such as UGC 10043 and LEDA 1656651, suggest that HCG~77 may be part of a more extended group that may be influencing its morphological and dynamical evolution.

We have identified several detailed structures and 16 distinct star-forming regions in the system using the high-resolution images. The photometry of these regions within the system was carried out, and their fluxes, luminosities, and sizes were derived. We carried out SED modeling to determine the physical properties of PGC~56121 and PGC~56125. Notably, PGC~56121 exhibits a higher stellar mass but a lower SFR and metallicity compared to PGC~56125. In contrast, PGC~56125 shows enhanced star formation activity and a more chemically enriched environment, indicating distinct evolutionary stages between the two galaxies.

\section*{Data availability}
L1 raw data from observations used in this paper are available at the AstroSat archives of ISRO maintained by ISSDC.\footnote{\url{https://astrobrowse.issdc.gov.in/astro_archive/archive/Home.jsp}} The final processed FUV image used for the analysis in this paper is available at the CDS via VizieR.\footnote{\url{http://cdsarc.u-strasbg.fr}}

\begin{acknowledgements}
      We gratefully acknowledge the use of data from the Ultra-Violet Imaging Telescope (UVIT) on board the {\it AstroSat} mission, an effort of the Indian Space Research Organisation (ISRO). The UVIT data used in this study were obtained through the data archive at the Indian Space Science Data Center (ISSDC). This research has used data from the NASA/IPAC Extragalactic Database (NED), which is funded by NASA and operated by the California Institute of Technology. This research has also used the Set of Identifications, Measurements, and Bibliographies for Astronomical Data (SIMBAD), which is maintained by the Centre de données astronomiques de Strasbourg (CDS).

The authors acknowledge the use of data from the Sloan Digital Sky Survey (SDSS). Funding for SDSS has been provided by the Alfred P. Sloan Foundation, the Participating Institutions, the National Science Foundation, and the U.S. Department of Energy Office of Science. This research has also used the Mikulski Archive for Space Telescopes (MAST). MAST is a NASA-funded archive maintained by the Space Telescope Science Institute (STScI), operated by the Association of Universities for Research in Astronomy, Inc., under NASA contract NAS 5–26555. This publication also used data products from the Wide field Infrared Survey Explorer (WISE), which is a joint project of the University of California, Los Angeles, and the Jet Propulsion Laboratory/California Institute of Technology, funded by the National Aeronautics and Space Administration.  K. P. Singh thanks the Indian National Science Academy for support under the Senior Scientist Programme. We thank Abhinna Sundar and Chavan Shende for their valuable contribution to this research work.
\end{acknowledgements}

\bibliography{aa}

@ARTICLE{1993ApL&C..29....1H,
       author = {{Hickson}, Paul},
        title = "{Atlas of Compact Groups of Galaxies (Special Issue)}",
      journal = {ApLC},
         year = 1993,
        month = jan,
       volume = {29},
        pages = {1},
       adsurl = {https://ui.adsabs.harvard.edu/abs/1993ApL\&C..29....1H}
}

@ARTICLE{1972ApJ...178..623T,
       author = {{Toomre}, Alar and {Toomre}, Juri},
        title = "{Galactic Bridges and Tails}",
      journal = {\apj},
         year = 1972,
        month = dec,
       volume = {178},
        pages = {623-666},
          doi = {10.1086/151823},
       adsurl = {https://ui.adsabs.harvard.edu/abs/1972ApJ...178..623T}
}

@ARTICLE{1992ARA&A..30..705B,
       author = {{Barnes}, Joshua E. and {Hernquist}, Lars},
        title = "{Dynamics of interacting galaxies.}",
      journal = {\araa},
         year = 1992,
        month = jan,
       volume = {30},
        pages = {705-742},
          doi = {10.1146/annurev.aa.30.090192.003421},
       adsurl = {https://ui.adsabs.harvard.edu/abs/1992ARA\&A..30..705B}
}

@ARTICLE{2021JApA...42...30P,
       author = {{Postma}, Joseph E. and {Leahy}, Denis},
        title = "{UVIT data reduction pipeline: A CCDLAB and UVIT tutorial}",
      journal = {JAA},
         year = 2021,
        month = oct,
       volume = {42},
       number = {2},
          eid = {30},
        pages = {30},
          doi = {10.1007/s12036-020-09689-w},
       adsurl = {https://ui.adsabs.harvard.edu/abs/2021JApA...42...30P}
}

@ARTICLE{1996A&AS..117..393B,
       author = {{Bertin}, E. and {Arnouts}, S.},
        title = "{SExtractor: Software for source extraction.}",
      journal = {\aaps},
         year = 1996,
        month = jun,
       volume = {117},
        pages = {393-404},
          doi = {10.1051/aas:1996164},
       adsurl = {https://ui.adsabs.harvard.edu/abs/1996A\&AS..117..393B}
}

@article{Tandon_2020,
   title={Additional Calibration of the Ultraviolet Imaging Telescope on Board AstroSat},
   volume={159},
   ISSN={1538-3881},
   url={http://dx.doi.org/10.3847/1538-3881/ab72a3},
   DOI={10.3847/1538-3881/ab72a3},
   number={4},
   journal={AJ},
   publisher={American Astronomical Society},
   author={Tandon, S. N. and Postma, J. and Joseph, P. and Devaraj, A. and Subramaniam, A and Barve, I. V. and George, K. and Ghosh, S. K. and Girish, V. and Hutchings, J. B. and Kamath, P. U. and Kathiravan, S. and Kumar, A. and Lancelot, J. P. and Leahy, D. and Mahesh, P. K. and Mohan, R. and Nagabhushana, S. and Pati, A. K. and Rao, N. Kameswara and Sankarasubramanian, K. and Sriram, S. and Stalin, C. S.},
   year={2020},
   month=mar, pages={158} }

@article{iglesias2006star,
  title={Star formation in the nearby universe: The ultraviolet and infrared points of view},
  author={Iglesias-Paramo, Jorge and Buat, V and Takeuchi, TT and Xu, K and Boissier, Samuel and Boselli, A and Burgarella, D and Madore, BF and De Paz, A Gil and Bianchi, L and others},
  journal={ApJS},
  volume={164},
  number={1},
  pages={38},
  year={2006},
  publisher={IOP Publishing}
}

@article{bianchi2011galex,
  title={GALEX and star formation},
  author={Bianchi, Luciana},
  journal={Ap\&SS},
  volume={335},
  pages={51--60},
  year={2011},
  publisher={Springer}
}

@article{schlafly2011measuring,
  title={Measuring reddening with Sloan Digital Sky Survey stellar spectra and recalibrating SFD},
  author={Schlafly, Edward F and Finkbeiner, Douglas P},
  journal={ApJ},
  volume={737},
  number={2},
  pages={103},
  year={2011},
  publisher={IOP Publishing}
}

@article{da_Cunha_2008,
   title={A simple model to interpret the ultraviolet, optical and infrared emission from galaxies},
   volume={388},
   ISSN={1365-2966},
   url={http://dx.doi.org/10.1111/j.1365-2966.2008.13535.x},
   DOI={10.1111/j.1365-2966.2008.13535.x},
   number={4},
   journal={MNRAS},
   publisher={Oxford University Press (OUP)},
   author={da Cunha, Elisabete and Charlot, Stéphane and Elbaz, David},
   year={2008},
   month=aug, pages={1595–1617} }

@article{Charlot_2000,
   title={A Simple Model for the Absorption of Starlight by Dust in Galaxies},
   volume={539},
   ISSN={1538-4357},
   url={http://dx.doi.org/10.1086/309250},
   DOI={10.1086/309250},
   number={2},
   journal={ApJ},
   publisher={American Astronomical Society},
   author={Charlot, Stephane and Fall, S. Michael},
   year={2000},
   month=aug, pages={718–731} }

@article{da_Cunha_2011,
   title={MAGPHYS: a publicly available tool to interpret observed galaxy SEDs},
   volume={7},
   ISSN={1743-9221},
   url={http://dx.doi.org/10.1017/S1743921312009283},
   DOI={10.1017/s1743921312009283},
   number={S284},
   journal={Proceedings of the International Astronomical Union},
   publisher={Cambridge University Press (CUP)},
   author={da Cunha, Elisabete and Charlot, Stéphane and Dunne, Loretta and Smith, Dan and Rowlands, Kate},
   year={2011},
   month=sep, pages={292–296} }

@ARTICLE{1982ApJ...255..382H,
       author = {{Hickson}, P.},
        title = "{Systematic properties of compact groups of galaxies.}",
      journal = {\apj},
         year = 1982,
        month = apr,
       volume = {255},
        pages = {382-391},
          doi = {10.1086/159838},
       adsurl = {https://ui.adsabs.harvard.edu/abs/1982ApJ...255..382H}
}

@article{Tandon_2017,
   title={In-orbit Calibrations of the Ultraviolet Imaging Telescope},
   volume={154},
   ISSN={1538-3881},
   url={http://dx.doi.org/10.3847/1538-3881/aa8451},
   DOI={10.3847/1538-3881/aa8451},
   number={3},
   journal={AJ},
   publisher={American Astronomical Society},
   author={Tandon, S. N. and Subramaniam, Annapurni and Girish, V. and Postma, J. and Sankarasubramanian, K. and Sriram, S. and Stalin, C. S. and Mondal, C. and Sahu, S. and Joseph, P. and Hutchings, J. and Ghosh, S. K. and Barve, I. V. and George, K. and Kamath, P. U. and Kathiravan, S. and Kumar, A. and Lancelot, J. P. and Leahy, D. and Mahesh, P. K. and Mohan, R. and Nagabhushana, S. and Pati, A. K. and Kameswara Rao, N. and Sreedhar, Y. H. and Sreekumar, P.},
   year={2017},
   month=sep, pages={128} }

@INPROCEEDINGS{2014SPIE.9144E..1SS,
       author = {{Singh}, Kulinder Pal and {Tandon}, S.~N. and {Agrawal}, P.~C. and {Antia}, H.~M. and {Manchanda}, R.~K. and {Yadav}, J.~S. and {Seetha}, S. and {Ramadevi}, M.~C. and {Rao}, A.~R. and {Bhattacharya}, D. and {Paul}, B. and {Sreekumar}, P. and {Bhattacharyya}, S. and {Stewart}, G.~C. and {Hutchings}, J. and {Annapurni}, S.~A. and {Ghosh}, S.~K. and {Murthy}, J. and {Pati}, A. and {Rao}, N.~K. and {Stalin}, C.~S. and {Girish}, V. and {Sankarasubramanian}, K. and {Vadawale}, S. and {Bhalerao}, V.~B. and {Dewangan}, G.~C. and {Dedhia}, D.~K. and {Hingar}, M.~K. and {Katoch}, T.~B. and {Kothare}, A.~T. and {Mirza}, I. and {Mukerjee}, K. and {Shah}, H. and {Shah}, P. and {Mohan}, R. and {Sangal}, A.~K. and {Nagabhusana}, S. and {Sriram}, S. and {Malkar}, J.~P. and {Sreekumar}, S. and {Abbey}, A.~F. and {Hansford}, G.~M. and {Beardmore}, A.~P. and {Sharma}, M.~R. and {Murthy}, S. and {Kulkarni}, R. and {Meena}, G. and {Babu}, V.~C. and {Postma}, J.},
        title = "{ASTROSAT mission}",
    booktitle = {Space Telescopes and Instrumentation 2014: Ultraviolet to Gamma Ray},
         year = 2014,
       editor = {{Takahashi}, Tadayuki and {den Herder}, Jan-Willem A. and {Bautz}, Mark},
       series = {SPIE Conference Series},
       volume = {9144},
        month = jul,
          eid = {91441S},
        pages = {91441S},
          doi = {10.1117/12.2062667},
       adsurl = {https://ui.adsabs.harvard.edu/abs/2014SPIE.9144E..1SS}
}

@article{Amrutha_2024,
   title={A comparative study of star-forming dwarf galaxies using the UVIT},
   volume={530},
   ISSN={1365-2966},
   url={http://dx.doi.org/10.1093/mnras/stae907},
   DOI={10.1093/mnras/stae907},
   number={2},
   journal={MNRAS},
   publisher={Oxford University Press (OUP)},
   author={Amrutha, S and Das, Mousumi and Yadav, Jyoti},
   year={2024},
   month=mar, pages={2199–2231} }

@ARTICLE{1996ApJ...464..641M,
       author = {{Mihos}, J. Christopher and {Hernquist}, Lars},
        title = "{Gasdynamics and Starbursts in Major Mergers}",
      journal = {\apj},
         year = 1996,
        month = jun,
       volume = {464},
        pages = {641},
          doi = {10.1086/177353},
archivePrefix = {arXiv},
       eprint = {astro-ph/9512099},
 primaryClass = {astro-ph},
       adsurl = {https://ui.adsabs.harvard.edu/abs/1996ApJ...464..641M}
}

@article{Hopkins_2008,
   title={A Cosmological Framework for the Co‐evolution of Quasars, Supermassive Black Holes, and Elliptical Galaxies. I. Galaxy Mergers and Quasar Activity},
   volume={175},
   ISSN={1538-4365},
   url={http://dx.doi.org/10.1086/524362},
   DOI={10.1086/524362},
   number={2},
   journal={ApJS},
   publisher={American Astronomical Society},
   author={Hopkins, Philip F. and Hernquist, Lars and Cox, Thomas J. and Kereš, Dušan},
   year={2008},
   month=apr, pages={356–389} }

@ARTICLE{1989Natur.340..687H,
       author = {{Hernquist}, Lars},
        title = "{Tidal triggering of starbursts and nuclear activity in galaxies}",
      journal = {\nat},
         year = 1989,
        month = aug,
       volume = {340},
       number = {6236},
        pages = {687-691},
          doi = {10.1038/340687a0},
       adsurl = {https://ui.adsabs.harvard.edu/abs/1989Natur.340..687H}
}

@article{Hopkins_2006,
   title={A Unified, Merger‐driven Model of the Origin of Starbursts, Quasars, the Cosmic X‐Ray Background, Supermassive Black Holes, and Galaxy Spheroids},
   volume={163},
   ISSN={1538-4365},
   url={http://dx.doi.org/10.1086/499298},
   DOI={10.1086/499298},
   number={1},
   journal={ApJS},
   publisher={American Astronomical Society},
   author={Hopkins, Philip F. and Hernquist, Lars and Cox, Thomas J. and Di Matteo, Tiziana and Robertson, Brant and Springel, Volker},
   year={2006},
   month=mar, pages={1–49} }

@article{Kormendy_2013,
   title={Coevolution (Or Not) of Supermassive Black Holes and Host Galaxies},
   volume={51},
   ISSN={1545-4282},
   url={http://dx.doi.org/10.1146/annurev-astro-082708-101811},
   DOI={10.1146/annurev-astro-082708-101811},
   number={1},
   journal={ARA\&A},
   publisher={Annual Reviews},
   author={Kormendy, John and Ho, Luis C.},
   year={2013},
   month=aug, pages={511–653} }

@article{10.1111/j.1365-2966.2008.13531.x,
    author = {McIntosh, Daniel H. and Guo, Yicheng and Hertzberg, Jen and Katz, Neal and Mo, H. J. and Van Den Bosch, Frank C. and Yang, Xiaohu},
    title = {Ongoing assembly of massive galaxies by major merging in large groups and clusters from the SDSS},
    journal = {MNRAS},
    volume = {388},
    number = {4},
    pages = {1537-1556},
    year = {2008},
    month = {08},
    issn = {0035-8711},
    doi = {10.1111/j.1365-2966.2008.13531.x},
    url = {https://doi.org/10.1111/j.1365-2966.2008.13531.x},
    eprint = {https://academic.oup.com/mnras/article-pdf/388/4/1537/3035119/mnras0388-1537.pdf},
}

@article{Samantaray_2024,
   title={AstroSat observations of interacting galaxies NGC 7469 and IC 5283},
   volume={686},
   ISSN={1432-0746},
   url={http://dx.doi.org/10.1051/0004-6361/202348981},
   DOI={10.1051/0004-6361/202348981},
   journal={A\&A},
   publisher={EDP Sciences},
   author={Samantaray, A. S. and Jassal, H. K. and Singh, K. P. and Dewangan, G. C.},
   year={2024},
   month=jun, pages={A241} }

@article{Mahajan_2022,
   title={Deepest far ultraviolet view of a central field in the Coma cluster byAstroSatUVIT},
   volume={39},
   ISSN={1448-6083},
   url={http://dx.doi.org/10.1017/pasa.2022.45},
   DOI={10.1017/pasa.2022.45},
   journal={PASA},
   publisher={Cambridge University Press (CUP)},
   author={Mahajan, Smriti and Singh, Kulinder Pal and Postma, Joseph E. and Pradeep, Kala G. and George, Koshy and Côté, Patrick},
   year={2022} }

@article{Wenger_2000,
   title={The SIMBAD astronomical database: The CDS reference database for astronomical objects},
   volume={143},
   ISSN={1286-4846},
   url={http://dx.doi.org/10.1051/aas:2000332},
   DOI={10.1051/aas:2000332},
   number={1},
   journal={A\&ASS},
   publisher={EDP Sciences},
   author={Wenger, M. and Ochsenbein, F. and Egret, D. and Dubois, P. and Bonnarel, F. and Borde, S. and Genova, F. and Jasniewicz, G. and Laloë, S. and Lesteven, S. and Monier, R.},
   year={2000},
   month=apr, pages={9–22} }

@INPROCEEDINGS{1991ASSL..171...89H,
       author = {{Helou}, G. and {Madore}, B.~F. and {Schmitz}, M. and {Bicay}, M.~D. and {Wu}, X. and {Bennett}, J.},
        title = "{The NASA/IPAC extragalactic database.}",
    booktitle = {Databases and On-line Data in Astronomy},
         year = 1991,
       editor = {{Albrecht}, M.~A. and {Egret}, D.},
       series = {ASSL},
       volume = {171},
        month = jan,
        pages = {89-106},
          doi = {10.1007/978-94-011-3250-3_10},
       adsurl = {https://ui.adsabs.harvard.edu/abs/1991ASSL..171...89H}
}

@article{Ahumada_2020,
   title={The 16th Data Release of the Sloan Digital Sky Surveys: First Release from the APOGEE-2 Southern Survey and Full Release of eBOSS Spectra},
   volume={249},
   ISSN={1538-4365},
   url={http://dx.doi.org/10.3847/1538-4365/ab929e},
   DOI={10.3847/1538-4365/ab929e},
   number={1},
   journal={ApJS},
   publisher={American Astronomical Society},
   author={Ahumada, Romina and Prieto, Carlos Allende and Almeida, Andrés and Anders, Friedrich and Anderson, Scott F. and Andrews, Brett H. and Anguiano, Borja and Arcodia, Riccardo and Armengaud, Eric and Aubert, Marie and Avila, Santiago and Avila-Reese, Vladimir and Badenes, Carles and Balland, Christophe and Barger, Kat and Barrera-Ballesteros, Jorge K. and Basu, Sarbani and Bautista, Julian and Beaton, Rachael L. and Beers, Timothy C. and Benavides, B. Izamar T. and Bender, Chad F. and Bernardi, Mariangela and Bershady, Matthew and Beutler, Florian and Bidin, Christian Moni and Bird, Jonathan and Bizyaev, Dmitry and Blanc, Guillermo A. and Blanton, Michael R. and Boquien, Médéric and Borissova, Jura and Bovy, Jo and Brandt, W. N. and Brinkmann, Jonathan and Brownstein, Joel R. and Bundy, Kevin and Bureau, Martin and Burgasser, Adam and Burtin, Etienne and Cano-Díaz, Mariana and Capasso, Raffaella and Cappellari, Michele and Carrera, Ricardo and Chabanier, Solène and Chaplin, William and Chapman, Michael and Cherinka, Brian and Chiappini, Cristina and Doohyun Choi, Peter and Chojnowski, S. Drew and Chung, Haeun and Clerc, Nicolas and Coffey, Damien and Comerford, Julia M. and Comparat, Johan and da Costa, Luiz and Cousinou, Marie-Claude and Covey, Kevin and Crane, Jeffrey D. and Cunha, Katia and Ilha, Gabriele da Silva and Dai 戴, Yu Sophia 昱 and Damsted, Sanna B. and Darling, Jeremy and Davidson, James W. and Davies, Roger and Dawson, Kyle and De, Nikhil and de la Macorra, Axel and De Lee, Nathan and Queiroz, Anna Bárbara de Andrade and Deconto Machado, Alice and de la Torre, Sylvain and Dell’Agli, Flavia and du Mas des Bourboux, Hélion and Diamond-Stanic, Aleksandar M. and Dillon, Sean and Donor, John and Drory, Niv and Duckworth, Chris and Dwelly, Tom and Ebelke, Garrett and Eftekharzadeh, Sarah and Davis Eigenbrot, Arthur and Elsworth, Yvonne P. and Eracleous, Mike and Erfanianfar, Ghazaleh and Escoffier, Stephanie and Fan, Xiaohui and Farr, Emily and Fernández-Trincado, José G. and Feuillet, Diane and Finoguenov, Alexis and Fofie, Patricia and Fraser-McKelvie, Amelia and Frinchaboy, Peter M. and Fromenteau, Sebastien and Fu, Hai and Galbany, Lluís and Garcia, Rafael A. and García-Hernández, D. A. and Oehmichen, Luis Alberto Garma and Ge, Junqiang and Maia, Marcio Antonio Geimba and Geisler, Doug and Gelfand, Joseph and Goddy, Julian and Gonzalez-Perez, Violeta and Grabowski, Kathleen and Green, Paul and Grier, Catherine J. and Guo, Hong and Guy, Julien and Harding, Paul and Hasselquist, Sten and Hawken, Adam James and Hayes, Christian R. and Hearty, Fred and Hekker, S. and Hogg, David W. and Holtzman, Jon A. and Horta, Danny and Hou, Jiamin and Hsieh, Bau-Ching and Huber, Daniel and Hunt, Jason A. S. and Chitham, J. Ider and Imig, Julie and Jaber, Mariana and Angel, Camilo Eduardo Jimenez and Johnson, Jennifer A. and Jones, Amy M. and Jönsson, Henrik and Jullo, Eric and Kim, Yerim and Kinemuchi, Karen and Kirkpatrick IV, Charles C. and Kite, George W. and Klaene, Mark and Kneib, Jean-Paul and Kollmeier, Juna A. and Kong, Hui and Kounkel, Marina and Krishnarao, Dhanesh and Lacerna, Ivan and Lan, Ting-Wen and Lane, Richard R. and Law, David R. and Le Goff, Jean-Marc and Leung, Henry W. and Lewis, Hannah and Li, Cheng and Lian, Jianhui and Lin 林俐, Lihwai 暉 and Long, Dan and Longa-Peña, Penélope and Lundgren, Britt and Lyke, Brad W. and Ted Mackereth, J. and MacLeod, Chelsea L. and Majewski, Steven R. and Manchado, Arturo and Maraston, Claudia and Martini, Paul and Masseron, Thomas and Masters 何凱, Karen L. 論 and Mathur, Savita and McDermid, Richard M. and Merloni, Andrea and Merrifield, Michael and Mészáros, Szabolcs and Miglio, Andrea and Minniti, Dante and Minsley, Rebecca and Miyaji, Takamitsu and Mohammad, Faizan Gohar and Mosser, Benoit and Mueller, Eva-Maria and Muna, Demitri and Muñoz-Gutiérrez, Andrea and Myers, Adam D. and Nadathur, Seshadri and Nair, Preethi and Nandra, Kirpal and do Nascimento, Janaina Correa and Nevin, Rebecca Jean and Newman, Jeffrey A. and Nidever, David L. and Nitschelm, Christian and Noterdaeme, Pasquier and O’Connell, Julia E. and Olmstead, Matthew D. and Oravetz, Daniel and Oravetz, Audrey and Osorio, Yeisson and Pace, Zachary J. and Padilla, Nelson and Palanque-Delabrouille, Nathalie and Palicio, Pedro A. and Pan, Hsi-An and Pan, Kaike and Parker, James and Paviot, Romain and Peirani, Sebastien and Ramŕez, Karla Peña and Penny, Samantha and Percival, Will J. and Perez-Fournon, Ismael and Pérez-Ràfols, Ignasi and Petitjean, Patrick and Pieri, Matthew M. and Pinsonneault, Marc and Poovelil, Vijith Jacob and Povick, Joshua Tyler and Prakash, Abhishek and Price-Whelan, Adrian M. and Raddick, M. Jordan and Raichoor, Anand and Ray, Amy and Rembold, Sandro Barboza and Rezaie, Mehdi and Riffel, Rogemar A. and Riffel, Rogério and Rix, Hans-Walter and Robin, Annie C. and Roman-Lopes, A. and Román-Zúñiga, Carlos and Rose, Benjamin and Ross, Ashley J. and Rossi, Graziano and Rowlands, Kate and Rubin, Kate H. R. and Salvato, Mara and Sánchez, Ariel G. and Sánchez-Menguiano, Laura and Sánchez-Gallego, José R. and Sayres, Conor and Schaefer, Adam and Schiavon, Ricardo P. and Schimoia, Jaderson S. and Schlafly, Edward and Schlegel, David and Schneider, Donald P. and Schultheis, Mathias and Schwope, Axel and Seo, Hee-Jong and Serenelli, Aldo and Shafieloo, Arman and Shamsi, Shoaib Jamal and Shao, Zhengyi and Shen, Shiyin and Shetrone, Matthew and Shirley, Raphael and Aguirre, Víctor Silva and Simon, Joshua D. and Skrutskie, M. F. and Slosar, Anže and Smethurst, Rebecca and Sobeck, Jennifer and Sodi, Bernardo Cervantes and Souto, Diogo and Stark, David V. and Stassun, Keivan G. and Steinmetz, Matthias and Stello, Dennis and Stermer, Julianna and Storchi-Bergmann, Thaisa and Streblyanska, Alina and Stringfellow, Guy S. and Stutz, Amelia and Suárez, Genaro and Sun, Jing and Taghizadeh-Popp, Manuchehr and Talbot, Michael S. and Tayar, Jamie and Thakar, Aniruddha R. and Theriault, Riley and Thomas, Daniel and Thomas, Zak C. and Tinker, Jeremy and Tojeiro, Rita and Toledo, Hector Hernandez and Tremonti, Christy A. and Troup, Nicholas W. and Tuttle, Sarah and Unda-Sanzana, Eduardo and Valentini, Marica and Vargas-González, Jaime and Vargas-Magaña, Mariana and Vázquez-Mata, Jose Antonio and Vivek, M. and Wake, David and Wang, Yuting and Weaver, Benjamin Alan and Weijmans, Anne-Marie and Wild, Vivienne and Wilson, John C. and Wilson, Robert F. and Wolthuis, Nathan and Wood-Vasey, W. M. and Yan, Renbin and Yang, Meng and Yèche, Christophe and Zamora, Olga and Zarrouk, Pauline and Zasowski, Gail and Zhang, Kai and Zhao, Cheng and Zhao, Gongbo and Zheng, Zheng and Zheng, Zheng and Zhu, Guangtun and Zou, Hu},
   year={2020},
   month=jun, pages={3} }

@ARTICLE{2020AJ....160..120J,
       author = {{J{\"o}nsson}, H. and {et al.}},
        title = "{APOGEE DR16 Synopsis}",
      journal = {\aj},
         year = 2020,
        month = aug,
       volume = {160},
       number = {3},
          eid = {120},
        pages = {120},
          doi = {10.3847/1538-3881/abae3c},
archivePrefix = {arXiv},
       eprint = {2007.09220},
 primaryClass = {astro-ph.SR},
       adsurl = {https://ui.adsabs.harvard.edu/abs/2020AJ....160..120J}
}

@INPROCEEDINGS{2009RMxAC..35..201A,
       author = {{Aguirre}, P. and {Uson}, J.~M. and {Matthews}, L.~D.},
        title = "{Discovery of a Small Group that Drives the Evolution of the Edge-On Spiral Galaxy UGC 10043}",
    booktitle = {Revista Mexicana de Astronomia y Astrofisica Conference Series},
         year = 2009,
       series = {RMxAC},
       volume = {35},
        month = may,
        pages = {201-202},
       adsurl = {https://ui.adsabs.harvard.edu/abs/2009RMxAC..35..201A}
}

@article{Matthews_2004,
doi = {10.1086/421363},
url = {https://dx.doi.org/10.1086/421363},
year = {2004},
month = {jul},
publisher = {},
volume = {128},
number = {1},
pages = {137},
author = {Matthews, L. D. and de Grijs, R.},
title = {Optical Imaging and Spectroscopy of the Edge-on Sbc Galaxy UGC 10043: Evidence for a Galactic Wind and a Peculiar Triaxial Bulge},
journal = {AJ}
}

@ARTICLE{1972ApJ...173...25S,
       author = {{Searle}, Leonard and {Sargent}, Wallace L.~W.},
        title = "{Inferences from the Composition of Two Dwarf Blue Galaxies}",
      journal = {\apj},
         year = 1972,
        month = apr,
       volume = {173},
        pages = {25},
          doi = {10.1086/151398},
       adsurl = {https://ui.adsabs.harvard.edu/abs/1972ApJ...173...25S}
}

@article{Kunth_2000,
   title={The most metal-poor galaxies},
   volume={10},
   ISSN={1432-0754},
   url={http://dx.doi.org/10.1007/s001590000005},
   DOI={10.1007/s001590000005},
   number={1–2},
   journal={A\&ARv},
   publisher={Springer Science and Business Media LLC},
   author={Kunth, D. and Östlin, G.},
   year={2000},
   month=jun, pages={1–79} }

@ARTICLE{1992ApJ...399..353H,
       author = {{Hickson}, Paul and {Mendes de Oliveira}, Claudia and {Huchra}, John P. and {Palumbo}, Giorgio G.},
        title = "{Dynamical Properties of Compact Groups of Galaxies}",
      journal = {\apj},
         year = 1992,
        month = nov,
       volume = {399},
        pages = {353},
          doi = {10.1086/171932},
       adsurl = {https://ui.adsabs.harvard.edu/abs/1992ApJ...399..353H}
}

@BOOK{1991rc3..book.....D,
       author = {{de Vaucouleurs}, Gerard and {de Vaucouleurs}, Antoinette and {Corwin}, Jr., Herold G. and {Buta}, Ronald J. and {Paturel}, Georges and {Fouque}, Pascal},
        title = "{Third Reference Catalogue of Bright Galaxies}",
         year = 1991,
       adsurl = {https://ui.adsabs.harvard.edu/abs/1991rc3..book.....D}
}

@ARTICLE{2005ApJS..160..149S,
       author = {{Springob}, Christopher M. and {Haynes}, Martha P. and {Giovanelli}, Riccardo and {Kent}, Brian R.},
        title = "{A Digital Archive of H I 21 Centimeter Line Spectra of Optically Targeted Galaxies}",
      journal = {\apjs},
         year = 2005,
        month = sep,
       volume = {160},
       number = {1},
        pages = {149-162},
          doi = {10.1086/431550},
archivePrefix = {arXiv},
       eprint = {astro-ph/0505025},
 primaryClass = {astro-ph},
       adsurl = {https://ui.adsabs.harvard.edu/abs/2005ApJS..160..149S}
}

@ARTICLE{2022ApJS..259...35A,
       author = {{Abdurro'uf} and {Accetta}, Katherine and {Aerts}, Conny and {Silva Aguirre}, V{\'\i}ctor and {Ahumada}, Romina and {Ajgaonkar}, Nikhil and {Filiz Ak}, N. and {Alam}, Shadab and {Allende Prieto}, Carlos and {Almeida}, Andr{\'e}s and {Anders}, Friedrich and {Anderson}, Scott F. and {Andrews}, Brett H. and {Anguiano}, Borja and {Aquino-Ort{\'\i}z}, Erik and {Arag{\'o}n-Salamanca}, Alfonso and {Argudo-Fern{\'a}ndez}, Maria and {Ata}, Metin and {Aubert}, Marie and {Avila-Reese}, Vladimir and {Badenes}, Carles and {Barb{\'a}}, Rodolfo H. and {Barger}, Kat and {Barrera-Ballesteros}, Jorge K. and {Beaton}, Rachael L. and {Beers}, Timothy C. and {Belfiore}, Francesco and {Bender}, Chad F. and {Bernardi}, Mariangela and {Bershady}, Matthew A. and {Beutler}, Florian and {Bidin}, Christian Moni and {Bird}, Jonathan C. and {Bizyaev}, Dmitry and {Blanc}, Guillermo A. and {Blanton}, Michael R. and {Boardman}, Nicholas Fraser and {Bolton}, Adam S. and {Boquien}, M{\'e}d{\'e}ric and {Borissova}, Jura and {Bovy}, Jo and {Brandt}, W.~N. and {Brown}, Jordan and {Brownstein}, Joel R. and {Brusa}, Marcella and {Buchner}, Johannes and {Bundy}, Kevin and {Burchett}, Joseph N. and {Bureau}, Martin and {Burgasser}, Adam and {Cabang}, Tuesday K. and {Campbell}, Stephanie and {Cappellari}, Michele and {Carlberg}, Joleen K. and {Wanderley}, F{\'a}bio Carneiro and {Carrera}, Ricardo and {Cash}, Jennifer and {Chen}, Yan-Ping and {Chen}, Wei-Huai and {Cherinka}, Brian and {Chiappini}, Cristina and {Choi}, Peter Doohyun and {Chojnowski}, S. Drew and {Chung}, Haeun and {Clerc}, Nicolas and {Cohen}, Roger E. and {Comerford}, Julia M. and {Comparat}, Johan and {da Costa}, Luiz and {Covey}, Kevin and {Crane}, Jeffrey D. and {Cruz-Gonzalez}, Irene and {Culhane}, Connor and {Cunha}, Katia and {Dai}, Y. Sophia and {Damke}, Guillermo and {Darling}, Jeremy and {Davidson}, Jr., James W. and {Davies}, Roger and {Dawson}, Kyle and {De Lee}, Nathan and {Diamond-Stanic}, Aleksandar M. and {Cano-D{\'\i}az}, Mariana and {S{\'a}nchez}, Helena Dom{\'\i}nguez and {Donor}, John and {Duckworth}, Chris and {Dwelly}, Tom and {Eisenstein}, Daniel J. and {Elsworth}, Yvonne P. and {Emsellem}, Eric and {Eracleous}, Mike and {Escoffier}, Stephanie and {Fan}, Xiaohui and {Farr}, Emily and {Feng}, Shuai and {Fern{\'a}ndez-Trincado}, Jos{\'e} G. and {Feuillet}, Diane and {Filipp}, Andreas and {Fillingham}, Sean P. and {Frinchaboy}, Peter M. and {Fromenteau}, Sebastien and {Galbany}, Llu{\'\i}s and {Garc{\'\i}a}, Rafael A. and {Garc{\'\i}a-Hern{\'a}ndez}, D.~A. and {Ge}, Junqiang and {Geisler}, Doug and {Gelfand}, Joseph and {G{\'e}ron}, Tobias and {Gibson}, Benjamin J. and {Goddy}, Julian and {Godoy-Rivera}, Diego and {Grabowski}, Kathleen and {Green}, Paul J. and {Greener}, Michael and {Grier}, Catherine J. and {Griffith}, Emily and {Guo}, Hong and {Guy}, Julien and {Hadjara}, Massinissa and {Harding}, Paul and {Hasselquist}, Sten and {Hayes}, Christian R. and {Hearty}, Fred and {Hern{\'a}ndez}, Jes{\'u}s and {Hill}, Lewis and {Hogg}, David W. and {Holtzman}, Jon A. and {Horta}, Danny and {Hsieh}, Bau-Ching and {Hsu}, Chin-Hao and {Hsu}, Yun-Hsin and {Huber}, Daniel and {Huertas-Company}, Marc and {Hutchinson}, Brian and {Hwang}, Ho Seong and {Ibarra-Medel}, H{\'e}ctor J. and {Chitham}, Jacob Ider and {Ilha}, Gabriele S. and {Imig}, Julie and {Jaekle}, Will and {Jayasinghe}, Tharindu and {Ji}, Xihan and {Johnson}, Jennifer A. and {Jones}, Amy and {J{\"o}nsson}, Henrik and {Katkov}, Ivan and {Khalatyan}, Dr., Arman and {Kinemuchi}, Karen and {Kisku}, Shobhit and {Knapen}, Johan H. and {Kneib}, Jean-Paul and {Kollmeier}, Juna A. and {Kong}, Miranda and {Kounkel}, Marina and {Kreckel}, Kathryn and {Krishnarao}, Dhanesh and {Lacerna}, Ivan and {Lane}, Richard R. and {Langgin}, Rachel and {Lavender}, Ramon and {Law}, David R. and {Lazarz}, Daniel and {Leung}, Henry W. and {Leung}, Ho-Hin and {Lewis}, Hannah M. and {Li}, Cheng and {Li}, Ran and {Lian}, Jianhui and {Liang}, Fu-Heng and {Lin}, Lihwai and {Lin}, Yen-Ting and {Lin}, Sicheng and {Lintott}, Chris and {Long}, Dan and {Longa-Pe{\~n}a}, Pen{\'e}lope and {L{\'o}pez-Cob{\'a}}, Carlos and {Lu}, Shengdong and {Lundgren}, Britt F. and {Luo}, Yuanze and {Mackereth}, J. Ted and {de la Macorra}, Axel and {Mahadevan}, Suvrath and {Majewski}, Steven R. and {Manchado}, Arturo and {Mandeville}, Travis and {Maraston}, Claudia and {Margalef-Bentabol}, Berta and {Masseron}, Thomas and {Masters}, Karen L. and {Mathur}, Savita and {McDermid}, Richard M. and {Mckay}, Myles and {Merloni}, Andrea and {Merrifield}, Michael and {Meszaros}, Szabolcs and {Miglio}, Andrea and {Di Mille}, Francesco and {Minniti}, Dante and {Minsley}, Rebecca and {Monachesi}, Antonela},
        title = "{The Seventeenth Data Release of the Sloan Digital Sky Surveys: Complete Release of MaNGA, MaStar, and APOGEE-2 Data}",
      journal = {\apjs},
         year = 2022,
        month = apr,
       volume = {259},
       number = {2},
          eid = {35},
        pages = {35},
          doi = {10.3847/1538-4365/ac4414},
archivePrefix = {arXiv},
       eprint = {2112.02026},
 primaryClass = {astro-ph.GA},
       adsurl = {https://ui.adsabs.harvard.edu/abs/2022ApJS..259...35A}
}

@article{Kaviraj_2011,
   title={Tidal dwarf galaxies in the nearby Universe: Tidal dwarf galaxies in the nearby Universe},
   volume={419},
   ISSN={0035-8711},
   url={http://dx.doi.org/10.1111/j.1365-2966.2011.19673.x},
   DOI={10.1111/j.1365-2966.2011.19673.x},
   number={1},
   journal={MNRAS},
   publisher={Oxford University Press (OUP)},
   author={Kaviraj, Sugata and Darg, Daniel and Lintott, Chris and Schawinski, Kevin and Silk, Joseph},
   year={2011},
   month=nov, pages={70–79} }

@ARTICLE{2017ApJS..233...25A,
       author = {{Albareti}, Franco D. and {Allende Prieto}, Carlos and {Almeida}, Andres and {Anders}, Friedrich and {Anderson}, Scott and {Andrews}, Brett H. and {Arag{\'o}n-Salamanca}, Alfonso and {Argudo-Fern{\'a}ndez}, Maria and {Armengaud}, Eric and {Aubourg}, Eric and {Avila-Reese}, Vladimir and {Badenes}, Carles and {Bailey}, Stephen and {Barbuy}, Beatriz and {Barger}, Kat and {Barrera-Ballesteros}, Jorge and {Bartosz}, Curtis and {Basu}, Sarbani and {Bates}, Dominic and {Battaglia}, Giuseppina and {Baumgarten}, Falk and {Baur}, Julien and {Bautista}, Julian and {Beers}, Timothy C. and {Belfiore}, Francesco and {Bershady}, Matthew and {Bertran de Lis}, Sara and {Bird}, Jonathan C. and {Bizyaev}, Dmitry and {Blanc}, Guillermo A. and {Blanton}, Michael and {Blomqvist}, Michael and {Bolton}, Adam S. and {Borissova}, J. and {Bovy}, Jo and {Brandt}, William Nielsen and {Brinkmann}, Jonathan and {Brownstein}, Joel R. and {Bundy}, Kevin and {Burtin}, Etienne and {Busca}, Nicol{\'a}s G. and {Camacho Chavez}, Hugo Orlando and {Cano D{\'\i}az}, M. and {Cappellari}, Michele and {Carrera}, Ricardo and {Chen}, Yanping and {Cherinka}, Brian and {Cheung}, Edmond and {Chiappini}, Cristina and {Chojnowski}, Drew and {Chuang}, Chia-Hsun and {Chung}, Haeun and {Cirolini}, Rafael Fernando and {Clerc}, Nicolas and {Cohen}, Roger E. and {Comerford}, Julia M. and {Comparat}, Johan and {Correa do Nascimento}, Janaina and {Cousinou}, Marie-Claude and {Covey}, Kevin and {Crane}, Jeffrey D. and {Croft}, Rupert and {Cunha}, Katia and {Darling}, Jeremy and {Davidson}, Jr., James W. and {Dawson}, Kyle and {Da Costa}, Luiz and {Da Silva Ilha}, Gabriele and {Deconto Machado}, Alice and {Delubac}, Timoth{\'e}e and {De Lee}, Nathan and {De la Macorra}, Axel and {De la Torre}, Sylvain and {Diamond-Stanic}, Aleksandar M. and {Donor}, John and {Downes}, Juan Jose and {Drory}, Niv and {Du}, Cheng and {Du Mas des Bourboux}, H{\'e}lion and {Dwelly}, Tom and {Ebelke}, Garrett and {Eigenbrot}, Arthur and {Eisenstein}, Daniel J. and {Elsworth}, Yvonne P. and {Emsellem}, Eric and {Eracleous}, Michael and {Escoffier}, Stephanie and {Evans}, Michael L. and {Falc{\'o}n-Barroso}, Jes{\'u}s and {Fan}, Xiaohui and {Favole}, Ginevra and {Fernandez-Alvar}, Emma and {Fernandez-Trincado}, J.~G. and {Feuillet}, Diane and {Fleming}, Scott W. and {Font-Ribera}, Andreu and {Freischlad}, Gordon and {Frinchaboy}, Peter and {Fu}, Hai and {Gao}, Yang and {Garcia}, Rafael A. and {Garcia-Dias}, R. and {Garcia-Hern{\'a}ndez}, D.~A. and {Garcia P{\'e}rez}, Ana E. and {Gaulme}, Patrick and {Ge}, Junqiang and {Geisler}, Douglas and {Gillespie}, Bruce and {Gil Marin}, Hector and {Girardi}, L{\'e}o and {Goddard}, Daniel and {Gomez Maqueo Chew}, Yilen and {Gonzalez-Perez}, Violeta and {Grabowski}, Kathleen and {Green}, Paul and {Grier}, Catherine J. and {Grier}, Thomas and {Guo}, Hong and {Guy}, Julien and {Hagen}, Alex and {Hall}, Matt and {Harding}, Paul and {Harley}, R.~E. and {Hasselquist}, Sten and {Hawley}, Suzanne and {Hayes}, Christian R. and {Hearty}, Fred and {Hekker}, Saskia and {Hernandez Toledo}, Hector and {Ho}, Shirley and {Hogg}, David W. and {Holley-Bockelmann}, Kelly and {Holtzman}, Jon A. and {Holzer}, Parker H. and {Hu}, Jian and {Huber}, Daniel and {Hutchinson}, Timothy Alan and {Hwang}, Ho Seong and {Ibarra-Medel}, H{\'e}ctor J. and {Ivans}, Inese I. and {Ivory}, KeShawn and {Jaehnig}, Kurt and {Jensen}, Trey W. and {Johnson}, Jennifer A. and {Jones}, Amy and {Jullo}, Eric and {Kallinger}, T. and {Kinemuchi}, Karen and {Kirkby}, David and {Klaene}, Mark and {Kneib}, Jean-Paul and {Kollmeier}, Juna A. and {Lacerna}, Ivan and {Lane}, Richard R. and {Lang}, Dustin and {Laurent}, Pierre and {Law}, David R. and {Leauthaud}, Alexie and {Le Goff}, Jean-Marc and {Li}, Chen and {Li}, Cheng and {Li}, Niu and {Li}, Ran and {Liang}, Fu-Heng and {Liang}, Yu and {Lima}, Marcos and {Lin}, Lihwai and {Lin}, Lin and {Lin}, Yen-Ting and {Liu}, Chao and {Long}, Dan and {Lucatello}, Sara and {MacDonald}, Nicholas and {MacLeod}, Chelsea L. and {Mackereth}, J. Ted and {Mahadevan}, Suvrath and {Maia}, Marcio Antonio Geimba and {Maiolino}, Roberto and {Majewski}, Steven R. and {Malanushenko}, Olena and {Malanushenko}, Viktor and {Mallmann}, N{\'\i}colas Dullius and {Manchado}, Arturo and {Maraston}, Claudia and {Marques-Chaves}, Rui and {Martinez Valpuesta}, Inma and {Masters}, Karen L. and {Mathur}, Savita and {McGreer}, Ian D. and {Merloni}, Andrea and {Merrifield}, Michael R. and {M{\'e}sz{\'a}ros}, Szabolcs and {Meza}, Andres and {Miglio}, Andrea and {Minchev}, Ivan and {Molaverdikhani}, Karan and {Montero-Dorta}, Antonio D. and {Mosser}, Benoit and {Muna}, Demitri and {Myers}, Adam},
        title = "{The 13th Data Release of the Sloan Digital Sky Survey: First Spectroscopic Data from the SDSS-IV Survey Mapping Nearby Galaxies at Apache Point Observatory}",
      journal = {\apjs},
         year = 2017,
        month = dec,
       volume = {233},
       number = {2},
          eid = {25},
        pages = {25},
          doi = {10.3847/1538-4365/aa8992},
archivePrefix = {arXiv},
       eprint = {1608.02013},
 primaryClass = {astro-ph.GA},
       adsurl = {https://ui.adsabs.harvard.edu/abs/2017ApJS..233...25A}
}

@article{Stierwalt_2015,
   title={TiNy TITANS: THE ROLE OF DWARF–DWARF INTERACTIONS IN LOW-MASS GALAXY EVOLUTION},
   volume={805},
   ISSN={1538-4357},
   url={http://dx.doi.org/10.1088/0004-637X/805/1/2},
   DOI={10.1088/0004-637x/805/1/2},
   number={1},
   journal={ApJ},
   publisher={American Astronomical Society},
   author={Stierwalt, S. and Besla, G. and Patton, D. and Johnson, K. and Kallivayalil, N. and Putman, M. and Privon, G. and Ross, G.},
   year={2015},
   month=may, pages={2} }

@article{ refId0,
	author = {{Subramanian, Smitha} and {Mondal, Chayan} and {Kalari, Venu}},
	title = {Effect of low-mass galaxy interactions on their star formation},
	DOI= "10.1051/0004-6361/202346536",
	url= "https://doi.org/10.1051/0004-6361/202346536",
	journal = {A\&A},
	year = 2024,
	volume = 681,
	pages = "A8",
}

@article{Martin_2020,
   title={The role of mergers and interactions in driving the evolution of dwarf galaxies over cosmic time},
   volume={500},
   ISSN={1365-2966},
   url={http://dx.doi.org/10.1093/mnras/staa3443},
   DOI={10.1093/mnras/staa3443},
   number={4},
   journal={MNRAS},
   publisher={Oxford University Press (OUP)},
   author={Martin, G and Jackson, R A and Kaviraj, S and Choi, H and Devriendt, J E G and Dubois, Y and Kimm, T and Kraljic, K and Peirani, S and Pichon, C and Volonteri, M and Yi, S K},
   year={2020},
   month=nov, pages={4937–4957} }

@article{Zaragoza_Cardiel_2024,
   title={Detection and characterization of detached tidal dwarf galaxies},
   volume={689},
   ISSN={1432-0746},
   url={http://dx.doi.org/10.1051/0004-6361/202450349},
   DOI={10.1051/0004-6361/202450349},
   journal={A\&A},
   publisher={EDP Sciences},
   author={Zaragoza-Cardiel, Javier and Smith, Beverly J. and Jones, Mark G. and Giroux, Mark L. and Toner, Shawn and Alzate, Jairo A. and Fernández-Arenas, David and Mayya, Divakara and Ortiz-León, Gisela and Portilla, Mauricio},
   year={2024},
   month=sep, pages={A206} }

@article{10.1093/mnras/stu1804,
    author = {Lelli, Federico and Verheijen, Marc and Fraternali, Filippo},
    title = {The triggering of starbursts in low-mass galaxies},
    journal = {MNRAS},
    volume = {445},
    number = {2},
    pages = {1694-1712},
    year = {2014},
    month = {10},
    issn = {0035-8711},
    doi = {10.1093/mnras/stu1804},
    url = {https://doi.org/10.1093/mnras/stu1804},
    eprint = {https://academic.oup.com/mnras/article-pdf/445/2/1694/18197528/stu1804.pdf},
}

@INPROCEEDINGS{2003ASPC..295..489J,
       author = {{Joye}, W.~A. and {Mandel}, E.},
        title = "{New Features of SAOImage DS9}",
    booktitle = {Astronomical Data Analysis Software and Systems XII},
         year = 2003,
       editor = {{Payne}, H.~E. and {Jedrzejewski}, R.~I. and {Hook}, R.~N.},
       series = {ASPCS},
       volume = {295},
        month = jan,
        pages = {489},
       adsurl = {https://ui.adsabs.harvard.edu/abs/2003ASPC..295..489J}
}

@article{Bahr_2025,
   title={UGC 10043 in depth: Dissecting the polar bulge and subtle low surface brightness features},
   volume={698},
   ISSN={1432-0746},
   url={http://dx.doi.org/10.1051/0004-6361/202554995},
   DOI={10.1051/0004-6361/202554995},
   journal={A\&A},
   publisher={EDP Sciences},
   author={Bahr, S. K. H. and Mosenkov, A. V.},
   year={2025},
   month=jun, pages={L21} }
\bibliographystyle{aa}
\end{document}